\documentclass{aa}  

\usepackage{graphicx}
\usepackage{txfonts}
\usepackage[dvipsnames]{xcolor}
\usepackage{booktabs} 
\usepackage{hyperref}
\hypersetup{
    colorlinks = true,
    linkcolor = blue,
    citecolor = blue,
    urlcolor = blue
}

\makeatother

\begin{document}

   \title{Brightest group and cluster galaxies as indicators of relaxation} 

   \author{K. Kelder
          \inst{1}\fnmsep\inst{2}\fnmsep\inst{3}
          \and
          P. Heinämäki
          \inst{1}
          \and
          P. Nurmi
          \inst{4}
          \and
          E. Tempel
          \inst{2}\fnmsep\inst{5}
          \and
          M. Einasto
          \inst{2}
          }

   \institute{Tuorla Observatory, Department of Physics and Astronomy, University of Turku, Vesilinnantie 5, 20014 Turku, Finland
        \and
    Tartu Observatory, University of Tartu, Observatooriumi 1, 61602 Tõravere, Estonia 
        \and
    Finnish Centre for Astronomy with ESO (FINCA), University of Turku, Vesilinnantie 5, 20014 Turku, Finland
        \and
    Department of Biodiversity Sciences, University of Turku, Vesilinnantie 5, 20014 Turku, Finland
        \and
    Estonian Academy of Sciences, Kohtu 6, 10130 Tallinn, Estonia
             }
   \date{Received 21 April 2026 / Accepted 11 August 2026}

% \abstract{}{}{}{}{} 
% 5 {} token are mandatory
 
  \abstract
  % CONTEXT heading (optional) leave it empty if necessary  
   {Galaxy groups and clusters are widely used to probe the evolution of the cosmic web and cosmology, while assuming that they are relaxed.}
  % 
  % AIMS heading (mandatory)
   {We identify the properties of the brightest halo galaxies (BHGs) that can be used to predict the most likely sample of dynamically relaxed host halos. Our work combines thoroughly studied galaxy clusters with less frequently analysed groups.} 
  %
  % METHODS heading (mandatory)
   {Our analysis was based on data from the IllustrisTNG simulations. We considered several observationally motivated parameters, including the offset of the BHG from the potential well of the host system ($d_\mathrm{off}$) and from the r-band luminosity centre ($d_\text{lum}$), the distance between the brightest and second-brightest galaxies ($d_{12}$), and the $r$-band magnitude gap between them ($\Delta m_{12}$). The primary analysis was performed at redshift $z=0$, with an additional investigation of the redshift evolution of halo relaxation up to $ z=1$. The observable proxies were applied to construct a halo mass function (HMF), which was then compared to the HMF of the relaxed sample defined from 3D information commonly used in theoretical approaches.}
  %
  % RESULTS heading (mandatory)
   {We find that $d_\mathrm{off}$ and $\Delta m_{12}$ are effective indicators of group and cluster relaxation, particularly when used in combination. The selection criteria of $d_\mathrm{off} < 0.05~R_{200}$ and $\Delta m_{12} > 1.6~\text{mag}$ allowed us to reproduce an HMF that closely matches that of the theoretically relaxed halo population. These criteria can be applied to observations up to $z\sim 0.2$ within a mass range $\text M_{200}\geq 10^{12.5}\text M_\odot$ ($\text M_{*\text{, BHG}}\gtrsim 10^{10.9}\text M_\odot$), including groups and clusters in the selection. In this mass range, 15 -- 23\% of the systems are considered fully relaxed at $z=0$. The fraction of relaxed haloes decreases with redshift up to $z\sim 0.4$, after which the decrease is far slower.}
  %
  % CONCLUSIONS heading (optional), leave it empty if necessary 
   {}


   \maketitle
   
%
%-------------------------------------------------------------------

%%%%%%%%%

\section{Introduction}

Groups and clusters of galaxies are among the most massive gravitationally bound structures in the Universe. They span a continuous mass interval from a few $10^{12}~\text{M}_\odot$ to an order of $10^{15}~\text{M}_\odot$, hosting roughly tens to several hundred galaxies within them. Due to their high masses, galaxy groups and especially clusters are essential for probing the dynamics and evolution of the cosmic web \citep{frenk_cold_1985}. Alongside analysis of the cosmic microwave background anisotropies, Ia supernovae, baryon acoustic oscillations, and weak gravitational lensing, galaxy clusters prove to be an excellent tool for constraining cosmological parameters, more specifically, the matter density $\Omega_m$ and the amplitude of fluctuations on the scale of $8~h ^{-1}~\text{Mpc}$ $\sigma_8$, via their mass function \citep{henry_x-ray_2009, vikhlinin_chandra_2009, mantz_observed_2010, schellenberger_hicosmo_2017, abdullah_constraining_2023, papageorgiou_cluster_2023, ghirardini_srgerosita_2024}.
The dynamical state of clusters (and groups) directly affects their visual appearance and morphology and also their halo formation time and concentration \citep[e.g.][and references therein]{neto_statistics_2007, 2010A&A...522A..92E, deluca_three_2021}.  
Cosmological models based on the cluster mass function (similarly to those based on X-rays, the Sunyaev-Zel'dovich effect, and dynamical methods) assume that the mass profiles of galaxy clusters are characteristic of \mbox{relaxed} systems. This implies that the systems are in dynamical equilibrium, virialised, and that their morphologies are relatively smooth (i.e. no strong evidence of substructuring). Relaxation is often considered the ultimate stage of group and cluster evolution. However, when a group or cluster has reached relaxation, it can still evolve, particularly during merger and accretion events, when interactions temporarily disrupt the entire system. When this occurs, the systems require a time on the order of a few billion years to re-establish relaxation \citep[e.g.][]{poole_impact_2006}. The inclusion of unrelaxed systems in dynamical-state-dependent analyses introduces biased constraints on cosmological parameters, potentially leading to incorrect conclusions about the dynamics and evolution of the large-scale structure of the Universe \citep[e.g.][]{voit_cosmevol_2005, old_galaxy_2018,pratt_galaxy_2019}. Robust criteria are therefore needed to distinguish between relaxed and unrelaxed systems. 

In theoretical models, cluster relaxation is often determined using three parameters \citep{neto_statistics_2007}: (1) the subhalo mass fraction \mbox{($f_\text{sub} = \text{M}_\text{sub, 200}/\text{M}_{200}$)}, (2) the centre-of-mass displacement \mbox{($s\equiv|\mathbf{r}_\text{p} -\mathbf{r}_\text{CM}|/r_\text{vir}$)}, and (3) virialisation ($2T/|U|$; see Sect. \ref{analysis_halo_relax} for a more detailed overview of the criteria). These parameters are well defined in cosmological simulations, where the full 6D phase space and total masses of the systems are accessible. 
Observational data, however, only provide two sky-projected spatial coordinates, the line-of-sight velocities, and dynamical masses. These are subject to selection effects and systematic biases. This observational limitation prevents a robust inference of the cluster dynamical states directly from these theoretical criteria, motivating the development of alternative observable diagnostics.

The properties of brightest group and cluster galaxies are some of the most promising observable diagnostics. They have been explored as proxies for the dynamical state of their hosts. In the following, we use the term brightest halo galaxy (BHG) to collectively refer to brightest cluster galaxies (BCGs) and brightest group galaxies (BGGs), and the term halo denotes galaxy groups and clusters, including their dark matter and baryonic components. Previous studies \citep[e.g.][]{lavoie_xxl_2016, lopes_optical_2018, casas_optical_2024} have advocated for using quantities such as the magnitude gap between the brightest and second brightest halo galaxies $\Delta m_{12}$ (or brightest and fourth brightest $\Delta m_{14}$) and the spatial separation between them $d_{12}$. Additionally, the offset of the BHG from the cluster X-ray or optical emission centre, peculiar velocities of the main galaxies, and substructuring of the halo have been used for similar applications \citep[e.g.][]{coziol_dynamical_2009, einasto_multimodality_2012}. 
These quantities reflect the distinct observational signatures of relaxed and unrelaxed systems. Systems that have experienced recent merger events (i.e. are unrelaxed) tend to exhibit large spatial offsets and peculiar velocities, numerous substructures, small magnitude gaps, and misalignments with the cosmic web orientation, whereas relaxed systems show smooth morphologies, cool cores, and centrally located BHGs with dominant luminosities \citep[e.g.][]{2010A&A...522A..92E, depropris_brightest_2020, einasto_galaxy_2024}.

\begin{figure}
    \centering
    \includegraphics[width=1.\linewidth]{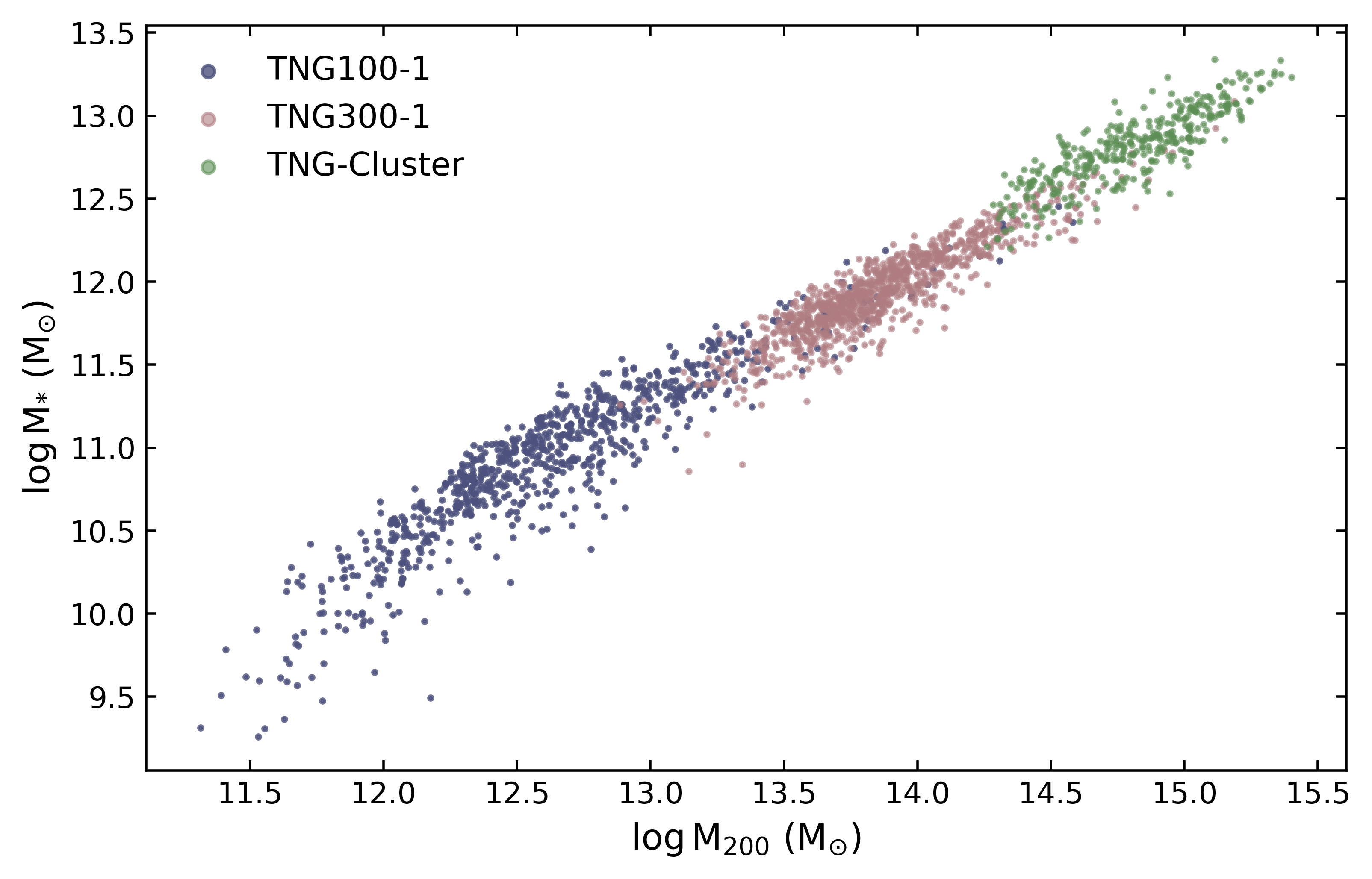}
    \caption{Relation of the BHG stellar mass to the virial mass of its host halo in the IllustrisTNG simulations at $z=0$. The colours represent different simulation runs: TNG100-1 (dark blue) covers the lower-mass end, TNG-Cluster (green) the massive end, and TNG300-1 (pink) the intermediate masses. The stellar masses of BHGs in the two larger simulations have been resolution corrected as described in Sect. \ref{resocorr_sect}} 
    \label{fig:BHG_vs_halo_200}
\end{figure}

The BHG-based diagnostics are feasible due to their exceptionally high luminosities and their unique formation history within their host systems. BHGs, which, according to the central galaxy paradigm \citep{van_den_bosch_CGP_2005}, are located near the centres of their host haloes, co-evolve with their host systems during hierarchical structure formation \citep[e.g.][]{ragone-figueroa_bcg_2018}. According to this framework, groups and clusters grow through bottom-up mass accumulation via mergers and accretion. During these interactions, a substantial amount of stellar mass is deposited near the centre of the dark matter halo and accreted by the BHG \citep[e.g.][]{kravtsov_borgani_formation_2012, cui_how_2016}. This evolutionary pathway establishes strong correlations between BHG properties and those of their host systems \citep{joeveer_spatial_1978, marini_veldisp_21, einasto_death_2022, sohn_illustristng_2022}.
The BHG stellar mass, in particular, serves as a robust proxy for estimating the virial mass of the host halo (see \mbox{Fig. \ref{fig:BHG_vs_halo_200}}). We typically observe that the BHG stellar mass roughly corresponds to 1\% of the total halo mass \citep[e.g.][]{lavoie_xxl_2016}. This relation, and others that connect BHGs to their host systems, have been studied thoroughly over the years in observational studies \citep[e.g.][]{oliva-altamirano_galaxy_2014, kravtsov_stellar_2018, erfanianfar_stellar_2019} and also in theoretical analyses, including semi-analytical models \citep[e.g.][]{de_lucia_hierarchical_2007} and hydrodynamical simulations \citep[e.g.][]{pillepich_first_2018}. These results confirmed that the properties of BHGs are strongly linked to their host systems. This validates their use as indicators.

We investigated how observable BHG properties can predict the dynamical state of their host systems across a wide range of group and cluster masses. While previous studies have primarily focused on massive clusters, we explicitly explored the full halo mass continuum to test the robustness of these diagnostics beyond the high-mass regime. We specifically focused on the offset of the BHG from the halo centre using two definitions for the halo centre and the separation and magnitude gap between the two brightest halo galaxies. Using data from the IllustrisTNG simulations, we examined how these observables relate to established theoretical relaxation criteria. We assessed their effectiveness as practical indicators of the halo dynamical state and the mass scales at which their predictive power begins to break down. 

The paper is organised as follows. In Sect. \ref{data} we provide an overview of the IllustrisTNG simulation and describe the data we used. Section \ref{analysis} outlines the analysis method. The results are presented in Sect. \ref{results}. Finally, in Sect. \ref{summary}, we discuss these results in the broader context of physical and cosmological implications.

\section{Data} \label{data}

\subsection{The IllustrisTNG simulations}

The data originate from the IllustrisTNG simulation suite \citep{nelson_first_2018, pillepich_first_2018, springel_first_2018, marinacci_first_2018, naiman_first_2018}. The TNG simulations consist of four simulation boxes – three in the main release \citep{nelson_illustristng_2018} are magnetohydrodynamical simulations, with box side lengths of 50, 100 and 300~Mpc, while the last, TNG-Cluster \citep{nelson_introducing_2023}, introduces a combination of 352 re-simulated cluster zoom magnetohydrodynamical simulations in a box with a side length of 1~Gpc. All simulations begin with cosmologically motivated initial conditions at redshift $z=127$, assuming a $\Lambda$CDM cosmology consistent with the \citet{planck_collaboration_planck_2016} results ($\Omega_{\Lambda,0} = 0.6911,\ 
\Omega_{m,0} = 0.3089,\ \Omega_{b,0} = 0.0486,$ \mbox{$\sigma_8 = 0.8159$}, $n_s = 0.9667$ and $h = 0.6774$).
The simulations are performed with the AREPO moving-mesh code \citep{springel_arepo_2010}, which is designed to model galaxy formation via a Voronoi tessellation, where the mesh points are allowed to move freely with the local fluid velocity. 
Further details on the physical models of galaxy formation implemented in IllustrisTNG, including treatments for star formation, feedback processes, and their numerical implementation, are presented in \citet{weinberger_simulating_2017} and \citet{pillepich_simulating_2018}.

The simulation group catalogues contain two types of objects: FoF haloes and Subfind subhaloes. The FoF haloes mainly refer to galaxy groups and clusters. They have been derived with the standard friends-of-friends (FoF) algorithm with a linking length $b=0.2$. The algorithm is run only on dark matter (DM) particles, while other particle types are later assigned to groups based on their nearest DM particle. The subhaloes (galaxies) are identified with the Subfind algorithm \citep{springel_subfind_2001}, where all particle types (gas, DM, stars) are considered in the calculations. The method first identifies all locally overdense regions within the FoF haloes as substructure candidates. The substructures are grown around particles with the highest density until all of them are assigned to a subgroup. Next, the boundedness of all particles is checked, and particles with a positive total energy are removed. When more than a threshold value of particles remain, they form a subhalo. For each FoF halo, Subfind returns the background halo, which is the largest object, mostly corresponding to the BHG, including intracluster light (ICL), and other subhaloes, which then correspond to satellite galaxies. Both the FoF haloes and Subfind subhaloes are available for each snapshot, covering 100 different redshifts. The catalogues are described thoroughly on the IllustrisTNG website's \footnote{\url{https://www.tng-project.org}} Data Specifications page.

\subsection{Data reductions}
To cover the full mass continuum of galaxy groups and clusters, the analysis includes data from the TNG100-1 and \mbox{TNG300-1} runs and from the TNG-Cluster high-resolution clusters. The baryon mass resolution in these simulations is $1.4 \cdot 10^6~\text{M}_\odot$, $1.1 \cdot 10^7~\text{M}_\odot$, and $1.2 \cdot 10^7~\text{M}_\odot$, respectively. The full mass range covered by these simulations is shown in Fig. \ref{fig:BHG_vs_halo_200}, where the halo virial mass ($\text{M}_{200}$) is shown as a function of the stellar mass of its brightest galaxy ($\text{M}_*$).  

Our selection of galaxies (subhaloes) was limited to objects with stellar mass $\text M_* \geq 10^9~\text M_\odot$, excluding lower-mass (lower-luminosity) systems because they contain far too few stellar particles to be sufficiently reliable in the simulations. We constrained the haloes by virial mass and richness (i.e. the number of member galaxies) in the main-release simulations, such that \mbox{TNG100-1} haloes had a mass of at least $10^{12.5}~\text{M}_\odot$ and contained a minimum of 3 members. In TNG300-1, since the resolution is lower, we chose haloes with at least 20 members, where their masses exceed $10^{12.8}~\text{M}_\odot$. All TNG-Cluster haloes contain at least 48 member galaxies (and are more massive than  $10^{14.2}~\text{M}_\odot$); therefore, no additional restrictions were made for this set.  To minimise possible boundary effects arising from the finite simulation volume (which may occur despite the use of periodic boundary conditions), we exclude haloes with centres located within 2 Mpc of the simulation box edges. 
Following this truncation, only subhaloes associated with haloes fully contained within the remaining volume are retained for the analysis. The resulting number of haloes and subhaloes at $z=0$ after these selections is reported in Table \ref{tab:data_used}. These data reduction steps were applied similarly to all simulation runs at redshifts $z = 0.2,\ 0.4,\ 0.6,\ 0.8,\ 1$. 

The brightest and second brightest members of these systems were chosen based on two criteria: the BHG must have an offset from the cluster potential well smaller than the halo virial radius $R_{200}~(\equiv R_{200c})$ and have the lowest r-band absolute magnitude. The absolute magnitudes are based on the summed luminosities of all stellar particles in the subhalo, as provided by the simulation. The distance is calculated between the centre of mass of the galaxy and the potential well of the halo. The selection of $d_\text{off}\leq R_{200}$ is a fairly loose criterion to ensure that the BHG can also be observationally identified as a halo member, while not imposing too strong constraints on the systems. This limit is especially important in large-volume simulations, where some systems may have offsets of a few megaparsec. 
The second brightest halo galaxy (SBHG) is not limited by distance and is selected as the brightest object in the r-band after the exclusion of the BHG. In the case of extensive merging systems, this can mean that some SBHGs, based solely on their magnitudes, might be slightly brighter than the BHG ($<5\%$ of systems). This type of phenomenon can also be observed in real galaxy groups and clusters. For instance, within the Fornax Cluster, the true brightest galaxy, NGC 1316, is located near the edge of the cluster, while the central galaxy, NGC 1399, is slightly fainter \citep{Drinkwater2001}. 
For consistency, in situations like these, we defined the BHG as the one closer to the halo's central region.

\begin{table}
    \caption{Size of the halo and subhalo samples in each simulation box after data reductions at $z=0$.}
    \label{tab:data_used}
    \centering
    \begin{tabular}{l r r}
    \hline\hline
    simulation run & halos & subhalos \\
    \hline
       TNG100-1    & 441 & 5838\\  
       TNG300-1    & 947 & 45658\\
       TNG-Cluster & 349 & 93515\\
    \hline
    \end{tabular}
\end{table}

\section{Analysis} \label{analysis}

The halo and BHG properties were investigated mainly at \mbox{$z=0$}, with separate redshift evolution analysis conducted, including redshifts $z = 0.2,\ 0.4,\ 0.6,\ 0.8,\ 1$. To simultaneously cover a broad range of system masses, from small groups to rich clusters, we combined data from the three IllustrisTNG simulation boxes. 

\subsection{Resolution corrections} \label{resocorr_sect}

\begin{figure}
    \centering
    \includegraphics[width=1.\linewidth]{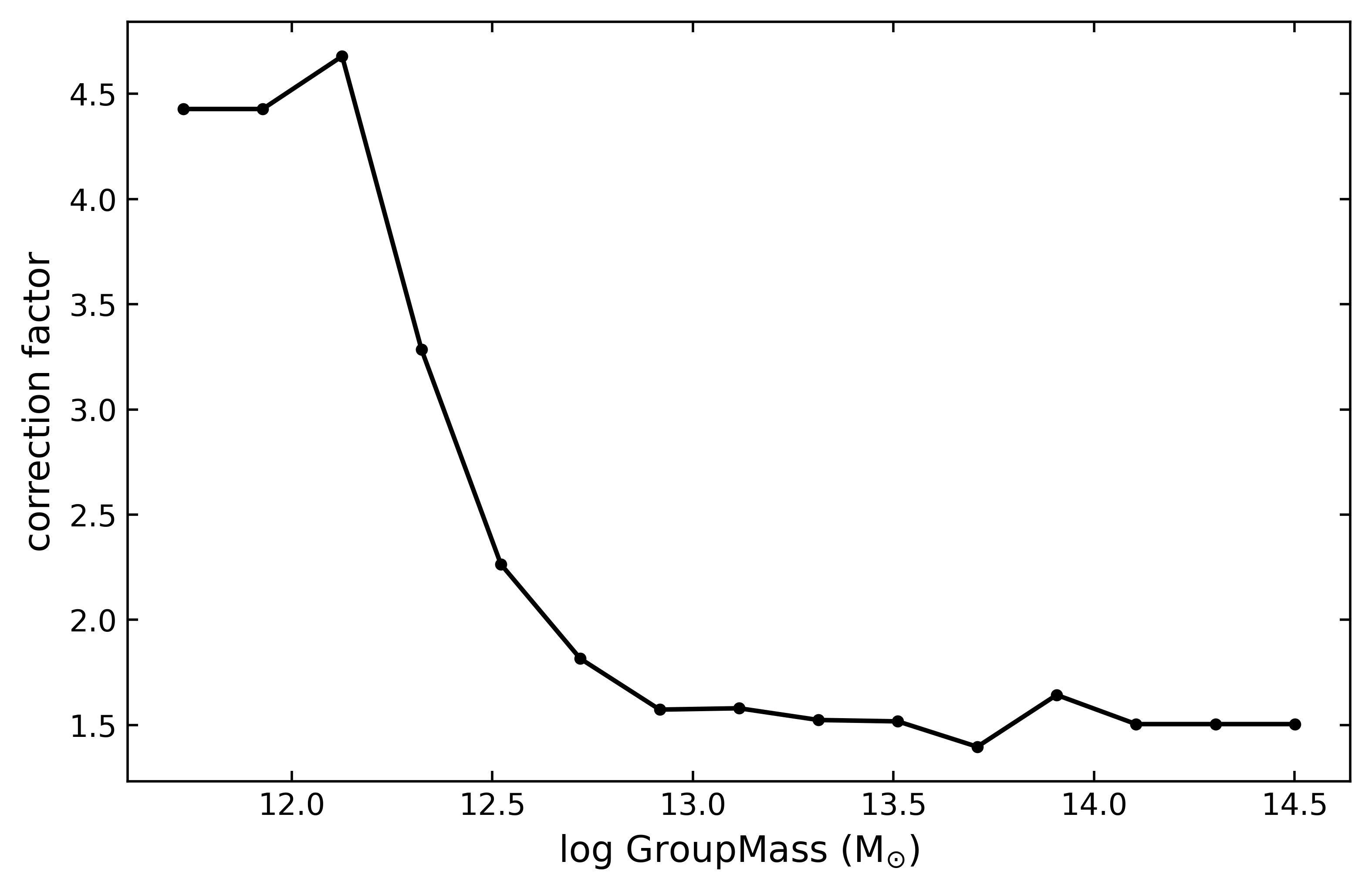}
    \caption{Resolution correction factors $\dfrac{\text{M}_*(\text{M}_\text{h}; \text{TNG100-1})}{\text{M}_*(\text{M}_\text{h}; \text{TNG100-2})}$ as a function of the total halo mass.}
    \label{fig:corr_factor_masses}
\end{figure}

Due to differences in resolution between the TNG100-1 simulation run compared with TNG300-1 and TNG-Cluster, it is necessary to correct for resolution effects to ensure that results of different simulation runs can be compared and combined. The main component of this work that fails to reach resolution convergence is the stellar mass ($\text M_*$) of the galaxies. To correct for this, we follow the discussion on simulation convergence by \citet{pillepich_first_2018}, who define the corrected stellar mass for TNG300-1 galaxies as
\begin{equation} \label{resocorr}
    \text{M}_*(\text{M}_\text{h}; \text{rTNG300}) = \text{M}_*(\text{M}_\text{h}; \text{TNG300}) \dfrac{\text{M}_*(\text{M}_\text{h}; \text{TNG100-1})}{\text{M}_*(\text{M}_\text{h}; \text{TNG100-2})},
\end{equation}
where $\text{M}_*(\text{M}_\text{h})$ is the sum of stellar masses for galaxies whose host haloes have their total mass (\texttt{GroupMass}) located in the corresponding mass bin $\text{M}_\text{h}$.
The fraction TNG100-1/TNG100-2 serves as a correction factor for different resolution levels, independent of volume effects. The correction is done in different \texttt{GroupMass} bins up to $\text{M}_\text{h} \lesssim 10^{14}\text{M}_\odot$. For halo masses greater than that, the correction factor is taken as an average across correction factors for bins $10^{13} \text{M}_\odot \leq \text{M}_\text{h} \leq 10^{14} \text{M}_\odot$. This averaging is necessary due to the small number of massive haloes in the TNG100 simulations.
Since the resolution of the TNG-Cluster haloes is roughly the same as for TNG300-1, the same correction factors were used for both data sets. The mass bins were defined as a range from the minimum to the maximum group mass, divided into 20 equal-width bins. Since the bins were fairly narrow ($\Delta \text{M} \sim 10^{0.2}~\text{M}_\odot$), the lowest-mass bin was often undersampled, making an accurate correction estimate difficult; therefore, the correction factor was extrapolated from the adjacent mass bins. The correction factors for each bin are displayed in Fig. \ref{fig:corr_factor_masses}, where it can be seen that the factor is high at low halo masses, indicating that at poorer resolution, we fail to fully resolve lower-mass systems, validating the choice of a lower mass limit for the haloes.

\subsection{Halo relaxation} \label{analysis_halo_relax}

We use the term relaxed halo to describe systems that are in dynamical equilibrium, virialised, and have a smooth morphology. To assess these criteria quantitatively, we apply the parameters and limits as introduced by \citet{neto_statistics_2007}, which are as follows: \mbox{(1) The} subhalo mass fraction $f_\text{sub} = \text{M}_\text{sub, 200}/\text{M}_{200} < 0.1$. Here $\text{M}_\text{sub, 200}$ accounts for the total mass of the subhaloes within the virial radius $R_{200}$. 
The total subhalo mass is calculated by taking a sum of the \texttt{SubhaloMass} values of each (non-BHG) galaxy that is located within $R_{200}$. The halo virial mass $\text M_{200}$ accounts for the total mass of all particle types enclosed in a sphere whose mean density is 200 times the critical density of the Universe. The chosen limit means that we assume that in the case of relaxation, less than 10\% of the halo mass is composed of satellite galaxies \citep[e.g.][]{springel_subfind_2001, de_lucia_hierarchical_2007, gouin_shape_2021}. 
\mbox{(2) The} centre of mass displacement $s = |\mathbf{r}_p - \mathbf{r}_\text{CM}|/R_{200} < 0.07$. The displacement criterion accounts for halo symmetry and presence of distinct substructures of subhaloes: when the halo centre of mass ($\mathbf{r}_\text{CM}$) is closely aligned with the gravitational potential well ($\mathbf{r}_{p}$), there is likely no ongoing interaction that would put the halo of interest out of equilibrium. We calculated the displacement as the 3D Euclidean distance using the centre of mass and potential well coordinates provided in the simulation data, computed from all particles bound to the halo.
\mbox{(3) The virial} ratio $2T/|U| < 1.35$. We adopted the method used by \citet{niemi_are_2007}, who compute the kinetic energy from the galaxies as $T = \sum_i m_iv_i^2/2$, where $v_i$ is the peculiar velocity of each galaxy within the virial radius, and $m_i$ is its mass. Likewise, the potential energy is calculated as $U = \sum_{i,j}Gm_i m_j/r_{i j}$, where $m_i,\ m_j$ are the masses of galaxies inside $R_{200}$ taken pairwise and $r_{i j}$ is the distance between them. 
In addition, the term for surface pressure energy has to be considered to account for the fact that haloes are not isolated systems. \citet{shaw_statistics_2006} introduce the virialisation criterion $\beta$, which includes surface pressure $E_S$ as $\beta = (2T-E_S)/U + 1$, where in the case of virialisation, the average value of $\beta$ should approximate to 0. The authors found that accounting for the surface pressure shifts the distribution of $\beta$ roughly 0.1 units towards more positive values. We applied the same shift to our criterion, which narrows the limit to $2T/|U|<1.25$. 

The dynamical state is assessed based on all three criteria simultaneously for method robustness. As these criteria can fluctuate during the virialisation process, we may fail to correctly identify which systems are briefly out of equilibrium. Since the fluctuations of the measures are assumed to be unsynchronised, it is unlikely that all three fail at the same time. A further, more detailed, discussion regarding the selection of the criteria can be found in papers by \citet{neto_statistics_2007, niemi_are_2007} and \citet{ludlow_dynamical_2012}.

\subsection{Observable proxies for relaxation} \label{analysis_obs_prox}

Once the set of relaxed haloes was obtained, we examined various properties of both haloes and their BHGs. Following the results found by \citet{lopes_optical_2018} and \citet{casas_optical_2024}, we delved into properties, such as the BHG offset from the halo centre $d_\text{off}$, distance between the two brightest galaxies $d_{12}$ and the magnitude gap between the BHG and SBHG $\Delta m_{12} = \text M_2 - \text M_1$. Regarding the definition of the halo centre, multiple approaches can be taken. The true halo centre, which corresponds to the potential well, is difficult to determine observationally, as we cannot directly observe dark matter. Gravitational lensing techniques provide tools for determining the total mass profiles for a large number of individual clusters. However, they are affected by various sources of statistical uncertainty and noise, making their interpretation feasible only for the more massive galaxy clusters 
\citep{euclid_collaboration_euclid_2024, euclid_collaboration_euclid_2025}.
In observations, the halo centre is often defined as the optical luminosity centre or centre of X-ray emission, as they are expected to lie near the potential well of the halo \citep[e.g.][]{seppi_offset_2023, popesso_perils_2025}. Since the IllustrisTNG suite lacks detailed X-ray emission information for most simulation runs, we analysed the BHG distances from the potential well and the luminosity centre. The potential well coordinates correspond to the particle with the lowest potential energy, and the luminosity centre is calculated as a luminosity-weighted "centre of mass" from the luminosities (derived from r-band absolute magnitudes) of all member galaxies.

The distances between the BHG and the halo potential well ($d_\text{off}$), the BHG and the halo luminosity centre ($d_\text{lum}$), and the BHG and the SBHG ($d_{12}$), are calculated as 3D Euclidean distances. We defined the centres of BHGs and SBHGs using their centre of mass, as provided by the simulation data in the \texttt{SubhaloCM} column. 
The magnitude gap between the two brightest galaxies has often been associated with the formation history of galaxy groups and clusters. While this property on its own is unable to trace the full merging and accretion history, it has been noted that large magnitude gaps suggest that the systems have formed early and there have been no recent major mergers, which can therefore imply relaxation \citep[e.g.][]{kundert_are_2017, vitorelli_mass_2018}.
We defined the magnitude gap $\Delta m_{12} = \text{M}_2 - \text{M}_1$ as the difference in r-band absolute magnitude. The absolute magnitudes are provided in the simulation data, where they are based on the summed luminosities of all stellar particles in the subhalo.

At $z=0$, the halo data from the three simulation runs were combined to achieve a greater coverage of the overlapping mass ranges. The data were split into virial mass bins between $10^{12.5}\text{M}_\odot$ and $10^{14.5}\text{M}_\odot$ with a step of $10^{0.5}\text{M}_\odot$ and one additional wider bin for the most massive systems. The number of haloes in each mass bin, starting with the lowest mass, is as follows: $[281,~203,~623,~293,~337]$. 
The fraction of relaxed haloes as a function of halo mass was examined in relation to several observable properties ($d_\text{off},\ d_\text{lum},\ d_{12},\ \Delta m_{12}$). To estimate the uncertainties in these results, we used a standard bootstrapping process with sample replacement. Given the varying number of data points in each distance and magnitude gap bin, we applied a Gaussian filter to the relaxed fraction values to smooth for these fluctuations. This is done using the Python \texttt{gaussian\_filter1d} package with $\sigma = \sqrt{2}$.

\section{Results} \label{results}

This section presents our results in three parts. First, we describe halo relaxation properties at redshift $z=0$. Second, we discuss the redshift evolution of these relations up to $z=1$. Finally, we examine whether the observable parameters can accurately reconstruct the relaxed halo mass function and assess how well this approach captures the true relaxed halo population.

\subsection{Halo relaxation at $z=0$}  \label{obs_prox_rel}

\begin{figure}
    \centering
    \includegraphics[width=1\linewidth]{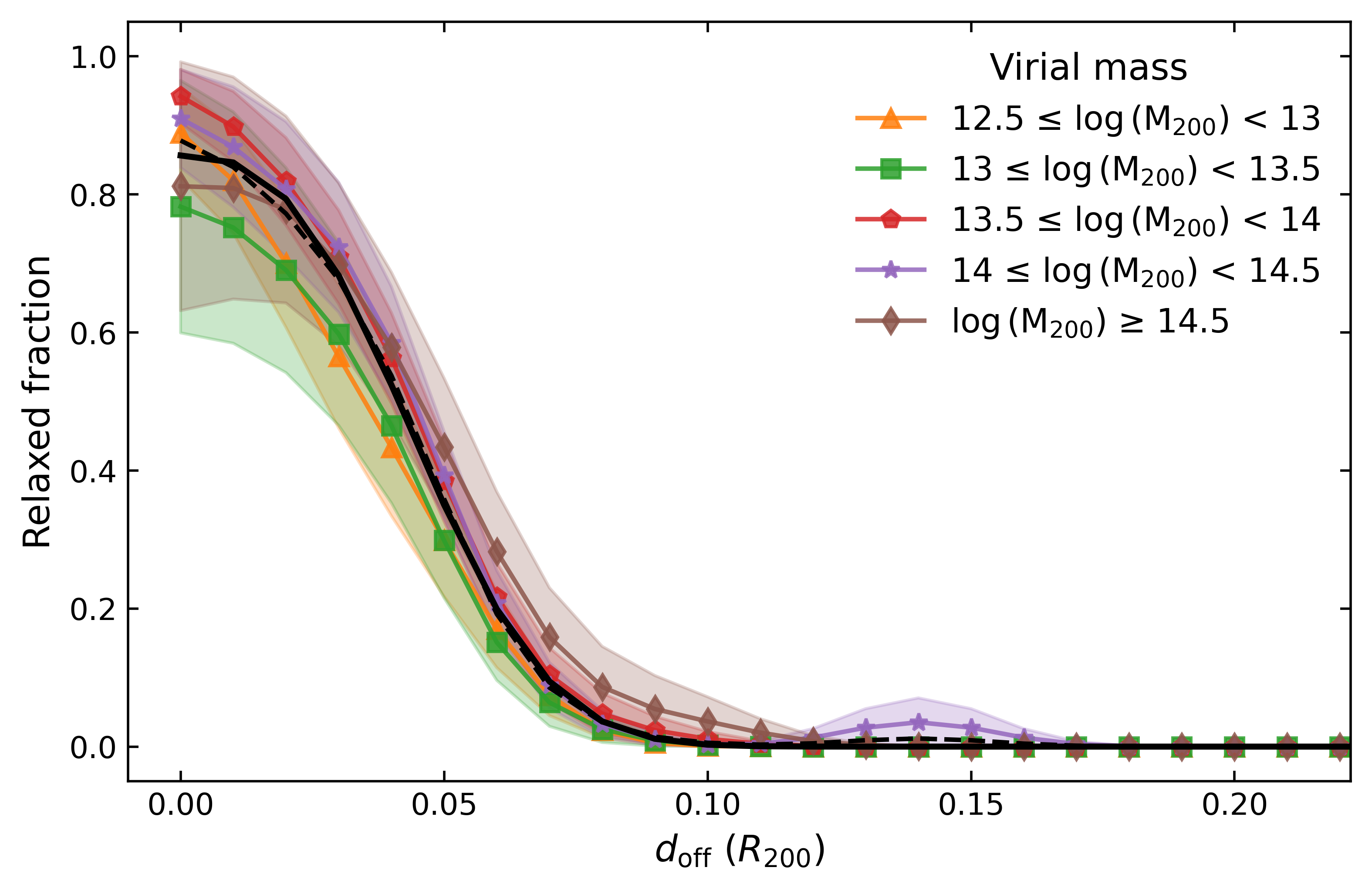}
    \caption{Fraction of relaxed haloes at different BHG offsets from the halo potential well (in $R_{200}$ units). The error bands indicate the 68\% bootstrapping confidence interval. The dashed black line represents the mean relaxed fraction, and the solid black line represents the best fit to a stretched exponential function (\ref{frel_doff}).}
    \label{fig:relfrac_doff}
\end{figure}

\begin{figure}
    \centering
    \includegraphics[width=1\linewidth]{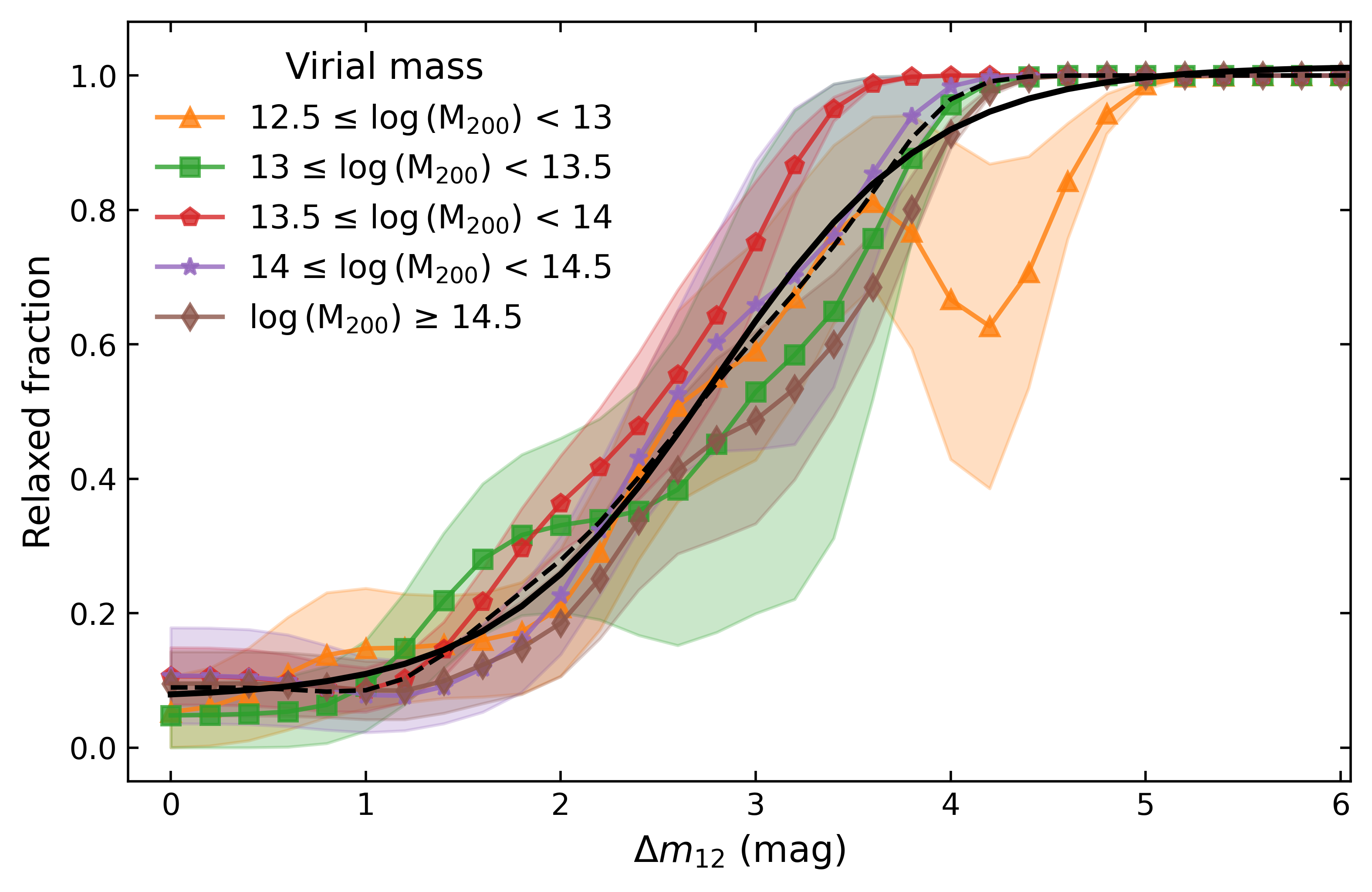}
    \caption{Fraction of relaxed haloes at different magnitude gap values. The error bands indicate the 68\% bootstrapping confidence interval. The dashed black line represents the weighted mean relaxed fraction, and the solid black line represents the best fit to a sigmoid function (\ref{frel_m12}).}
    \label{fig:relfrac_mgap}
\end{figure}

\begin{figure}
    \centering
    \includegraphics[width=1.\linewidth]{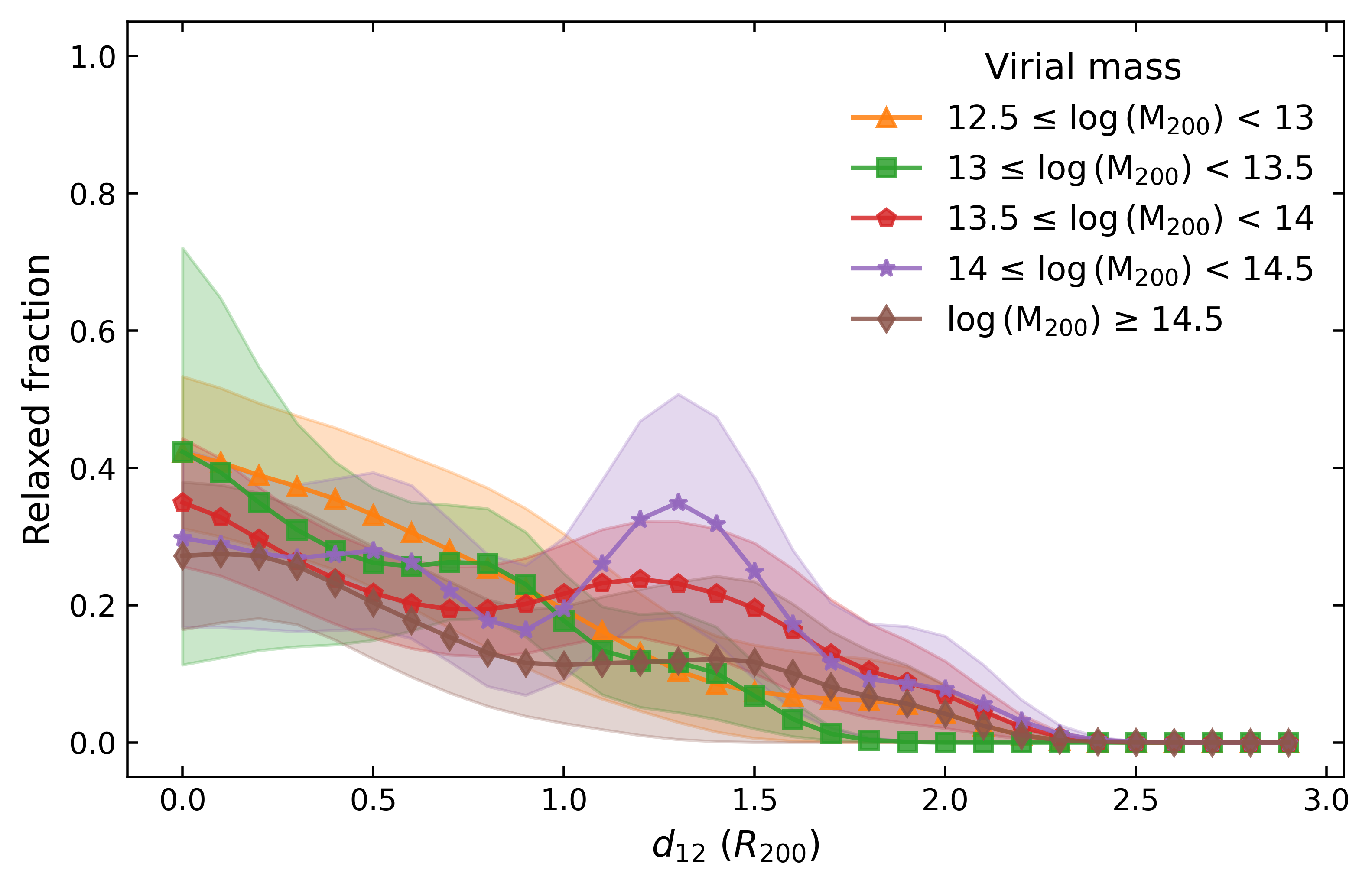}
    \caption{Fraction of relaxed haloes at different BHG -- SBHG distances (in $R_{200}$ units). The error bands indicate the 68\% bootstrapping confidence interval.} 
    \label{fig:relfrac_d12}
\end{figure}

\begin{figure}
    \centering
    \includegraphics[width=1.\linewidth]{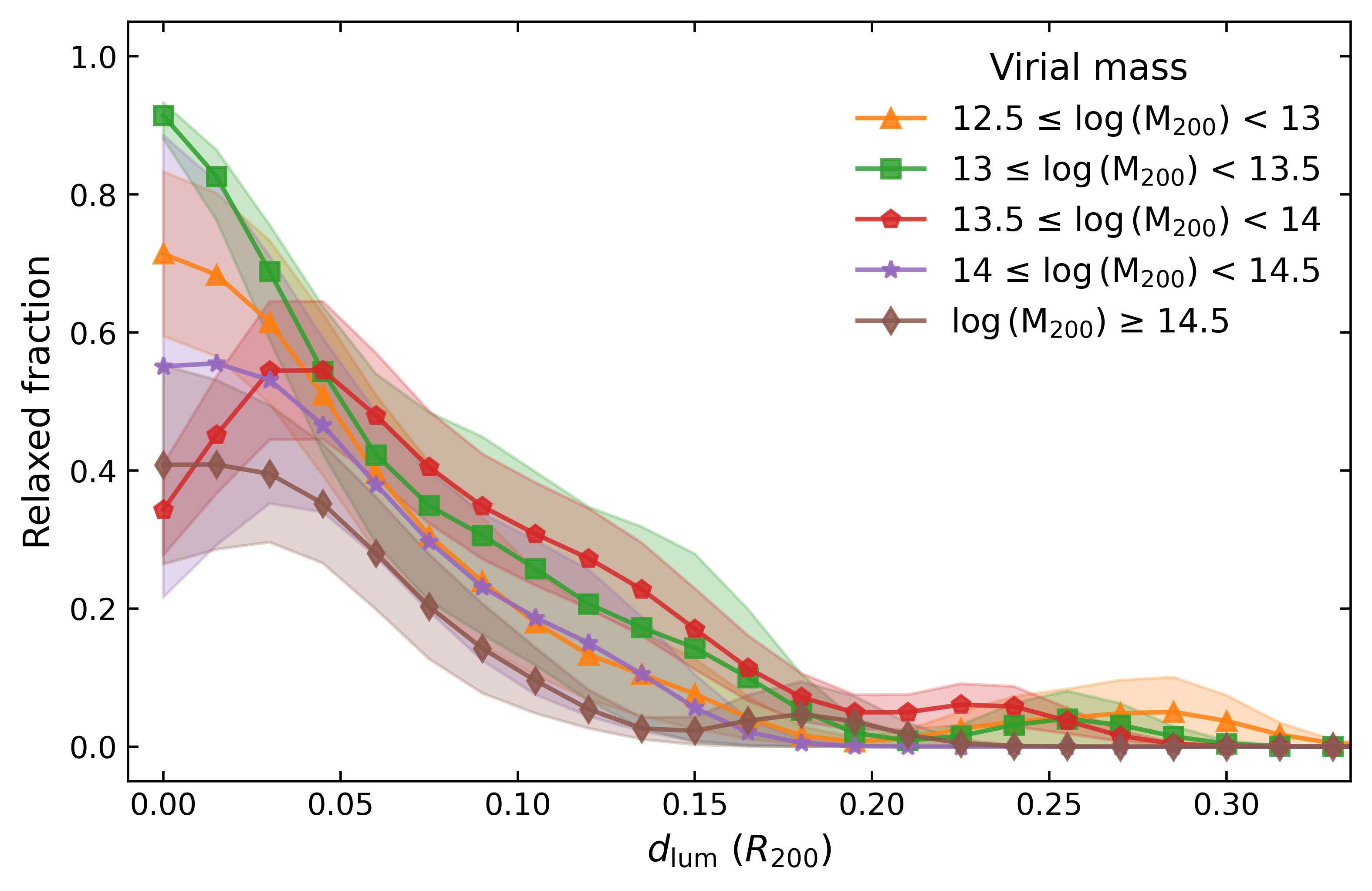}
    \caption{Fraction of relaxed haloes at different BHG offsets from the halo luminosity centre (in $R_{200}$ units). Error bands indicate a 68\% bootstrapping confidence interval. This metric performs best at lower halo masses.}
    \label{fig:relaxed_dlum}
\end{figure}

Figures \ref{fig:relfrac_doff} and \ref{fig:relfrac_mgap} display the relation between the fraction of relaxed haloes as a function of the BHG offset from the halo potential well ($d_\text{off}$) and as a function of the r-band absolute magnitude gap between the two brightest member galaxies ($\Delta m_{12}$). The distance between the two brightest galaxies ($d_{12}$) showed no strong correlation with halo relaxation (see Fig. \ref{fig:relfrac_d12}). While previous studies, such as \citet{casas_optical_2024}, found that relaxed haloes exhibit large $d_{12}$ values, our larger sample indicates that this parameter, while showing a weak anticorrelation with distance, is unreliable for estimating halo relaxation. Hence, the relation $d_{12}$ will be disregarded in the further analysis. The distance of the BHG from the luminosity centre ($d_\text{lum}$) is shown in Fig. \ref{fig:relaxed_dlum}. It proves to be a stronger indicator for relaxation than $d_{12}$, especially for lower mass systems $12.5 \leq \log \text{M}_{200} < 13.5$. In more massive systems, there is either a dip or a cut-off in the relaxed fraction at low offset values. This could imply that massive clusters have greater disruption in their stellar components, leading the luminosity centre to be an insufficient tracer of the cluster centre. Since this parameter performs worse in the full mass range than $d_\text{off}$, we will continue with the latter for finding a proxy of relaxation.

The relations are expressed in various halo virial mass bins as described in the previous section. Our sample contains 15 -- 23\% of relaxed systems, where the exact fraction depends on the halo mass (see $z=0$ fractions in Fig.\ref{fig:relfrac_massbin}).
This fraction agrees well with \citet[20\%]{einasto_multimodality_2012}, but is lower than what is found from some other studies of rich galaxy clusters with various definitions of relaxation. For example, in the Sloan Digital Sky Survey (SDSS) photometric data at $0.05 < z \lesssim 0.42$, the relaxation fraction was found to be 28\% \citep{wen_substructure_2013}, in the eROSITA X-ray data at $0.15 < z < 0.4$, it was 31\% \citep{seppi_offset_2023}, while at low redshifts ($0.017 < z < 0.02$) the observed fraction reached roughly $50\%$ \citep{ghirardini_erosita_2022}. This difference indicates that, in observations, we are more inclined to detect relaxed systems due to observational biases. This can also depend on the type of observations as well as the definition of relaxation; for instance, \citet{rossetti_cool-core_2017} showed that clusters detected with the Sunyaev–Zeldovich effect show a much lower fraction of cool-core clusters ($29 \pm 4\%$) than X-ray-selected samples ($59\pm 5\%$). They note that while X-ray-selected samples are known to be biased towards cool cores due to their prominent surface brightness peaks, the details of the selection effects that cause these biases remain to be investigated. A detailed overview of relaxation fractions obtained from different observational studies is provided by \citet[see their Table 10 for a summary of relaxation fraction measurements for various cluster samples]{gassis_chandra_2026}.

The BHG offset from the halo potential well is expressed in halo virial radius $R_{200}$ units to account for the dependence on halo physical size. We observed that more massive groups and clusters can accommodate larger BHG offsets while still maintaining relaxation. 
For a halo to be considered relaxed, its BHG must be located at least within the inner $10\%$ of the halo's virial radius. To quantify the relation between the BHG offset and the halo relaxation within this inner $10\%$, we calculated the mean curve over all mass bins and fit it with a stretched exponential function,
\begin{equation} \label{frel_doff}
    f_\text{rel}(d_\text{off}) = N_0\exp[(-\lambda~d_\text{off})^\beta].
\end{equation}
Here, $N_0$ is the amplitude, $\lambda$ is the characteristic length scale, and the factor $\beta$ expresses the stretch ($\beta < 1$) or compression ($\beta > 1$) of the exponential function. The best fit results yielded the following values: $N_0 = (0.857 \pm 0.003)$, $\lambda = (19.606 \pm 0.073)$, and $\beta = (2.493 \pm 0.033)$. This curve is shown with a solid black line in Fig. \ref{fig:relfrac_doff}. 
For the $\Delta m_{12}$ relation in Fig. \ref{fig:relfrac_mgap}, it must be stressed that the number of haloes in the high magnitude gap bins \mbox{(5 -- 6~mag)} is small, often only 1, meaning there can be no reliable bootstrapping uncertainties obtained from that range. Such systems are also very rare in observation \citep{Jones2003}.
Since all curves end up at a relaxed fraction of 1 at larger magnitude gaps, we again found the mean curve over the mass bins. Due to undersampling at higher $\Delta m_{12}$ values in the lower-mass bins, we calculated the curve as a weighted mean, with weights equal to the number of data points in each bin. The behaviour of this curve resembles a sigmoid function,
\begin{equation} \label{frel_m12}
    f_\text{rel}(\Delta m_{12}) = \dfrac{L}{1+e^{-x}} + b,
\end{equation}
where $x = k~(\Delta m_{12} - x_0)$, $\ L$ is the upper asymptote or maximum of the function (in the standard case $L=1$), $k$ is the logistic growth rate, that is, the steepness of the curve, and $x_{0}$ is the $x$ value of the function's midpoint (here, the midpoint of $\Delta m_{12}$) and $b$ is the offset along the y-axis. 
The best-fit values are the following: $L = (0.942 \pm 0.006)$, $x_0 = (2.772 \pm 0.012)$, \mbox{$k =  (1.684 \pm 0.032)$}, and $b =  (0.065 \pm 0.005)$.

\begin{figure}
    \centering
    \includegraphics[width=1.\linewidth]{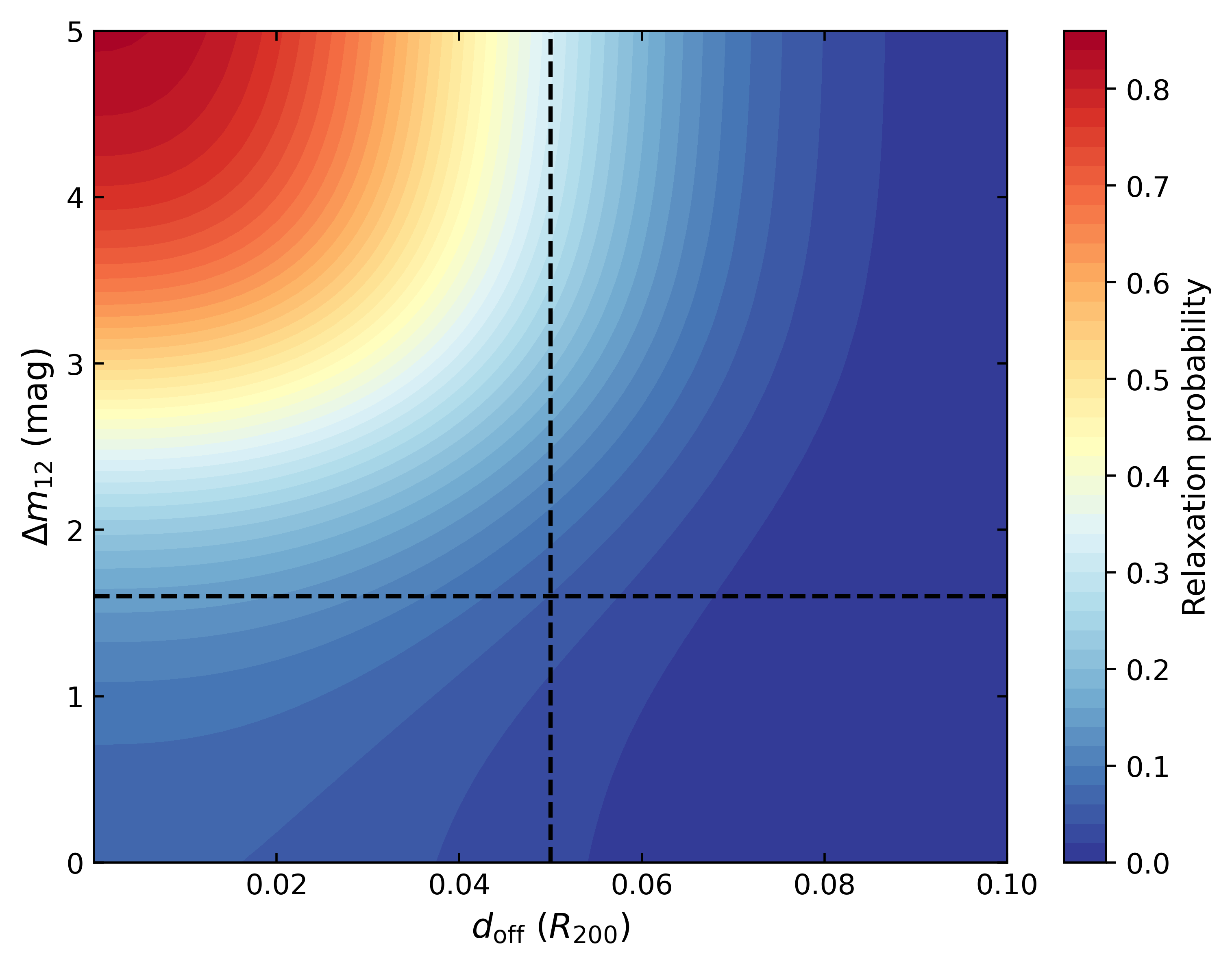}
    \caption{Relaxation probability of haloes in a 2D parameter space where the total fraction is a product of the two fitted functions (\ref{frel_doff}, \ref{frel_m12}) at each grid point value. The black lines correspond to limits in \mbox{Figs. \ref{fig:obslims}, \ref{fig:relfrac_from_data_3D2D}}.}
    \label{fig:relfrac_combined_multip}
\end{figure}

These results are useful initial observationally derived estimates for the dynamical state of the haloes. When we lack further details about these systems, we can make approximations about their relaxation based on these relations. The fitted functions (\ref{frel_doff}) and (\ref{frel_m12}) can also be combined in a 2D parameter space to obtain a more robust pairwise estimate. In other words, we obtain a 2D probability distribution of a halo being relaxed, given that it has certain BHG offset and magnitude gap values. The probability distribution is shown in Fig. \ref{fig:relfrac_combined_multip}. Here, the joint probability is taken as the product of the functions evaluated at each parameter-pair grid point. 
It is important to acknowledge that this assessment of the joint probability assumes that the two properties are completely independent at values $d_\text{off} \leq 0.1R_{200}$ and $\Delta m_{12} \leq 5~\text{mag}$. As a first-order approximation, we can assume that this is sufficient, but these properties evolve in parallel with the halo evolution (see Sect. \ref{redshift_evol}), so this relation might over- or underestimate the probabilities. 

\begin{figure}
    \centering
    \includegraphics[width=1.\linewidth]{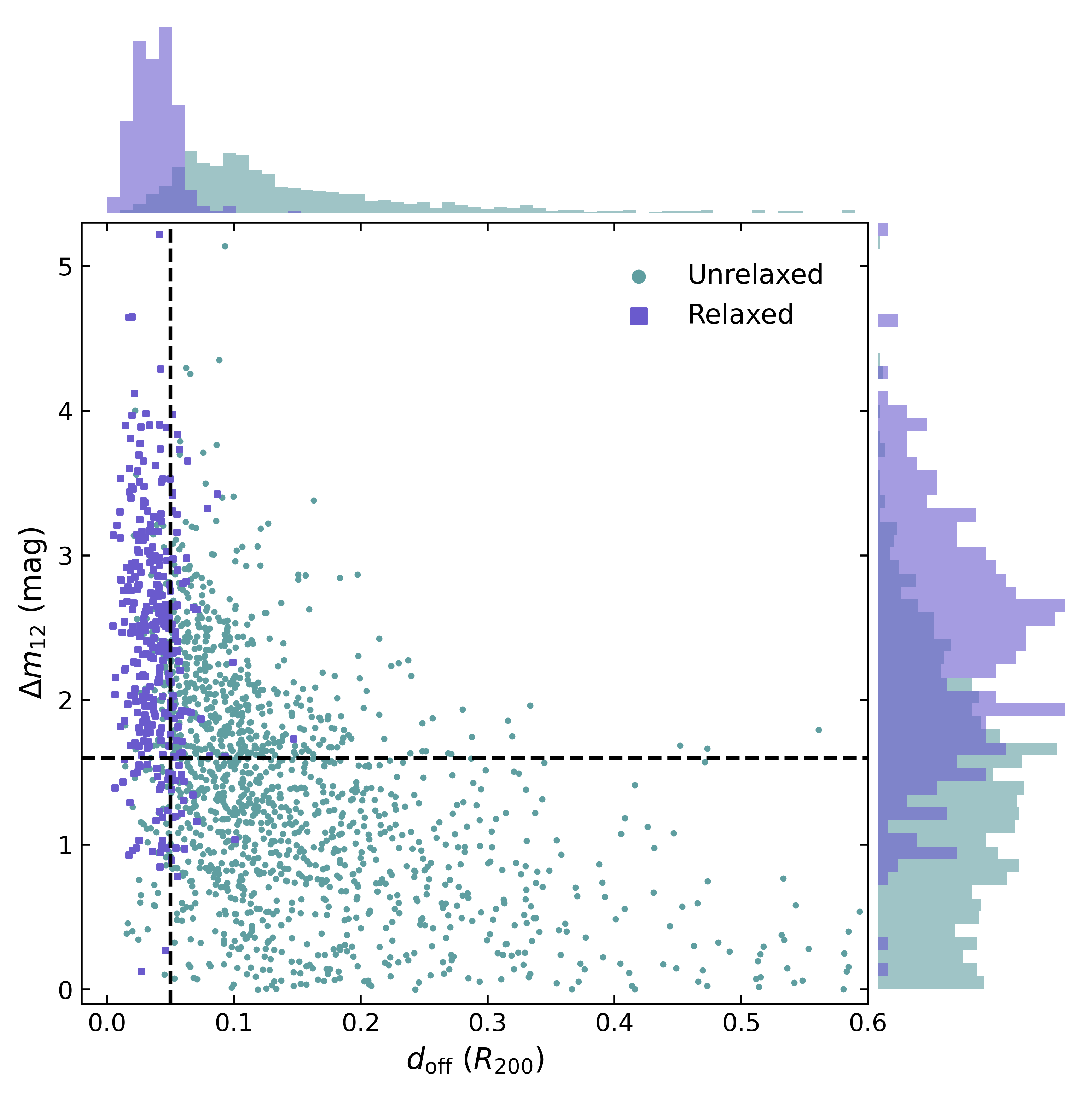}
    \caption{Parameter space of the offset of the BHG from the halo potential well and the magnitude gap for relaxed and unrelaxed haloes (defined in Sect. \ref{analysis_halo_relax}). The dashed black lines constrain a region with an 80\% relaxed fraction. The limits applied here are: $\Delta m_{12} > 1.6~\text{mag}$, \mbox{$d_\text{off} < 0.05~R_{200}$.} The histograms depict the halo number densities at different axis values.}
    \label{fig:obslims}
\end{figure}

\begin{figure*}
    \centering
    \includegraphics[width=.95\linewidth]{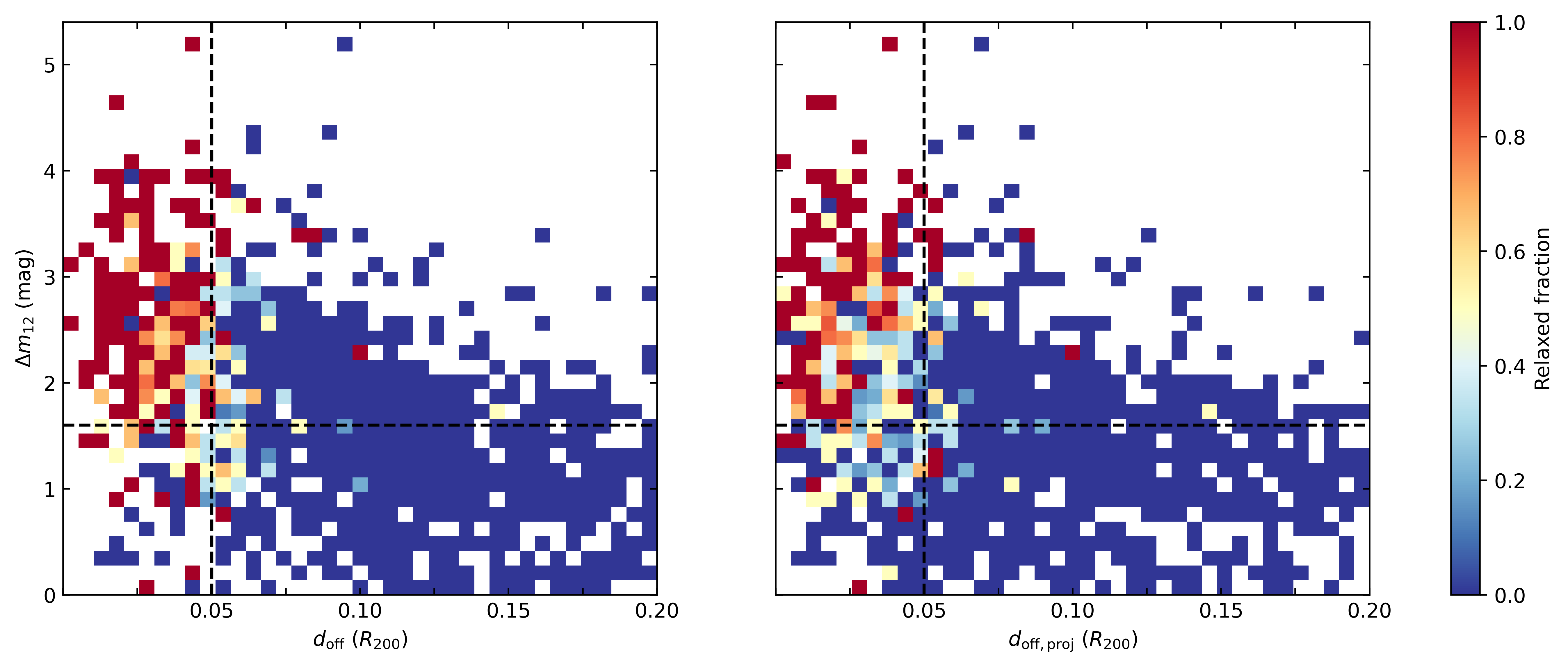}
    \caption{Relaxation fraction based on a direct sampling onto a grid from the data. \textit{Left}: Fractions for true 3D offsets $d_\text{off}$ from the halo centre. \textit{Right}: Fractions for the projected 2D offsets $d_\text{off, proj}$. The dashed black lines indicate limits for the proxy \mbox{sample:} $\Delta m_{12} > 1.6~\text{mag}$, $d_\text{off} < 0.05~R_{200}.$ The offset limit is a suitable constraint in both the 3D and 2D cases.}
    \label{fig:relfrac_from_data_3D2D}
\end{figure*}

Another way to assess the dynamical state with the two parameters is to directly determine reliable limits from the distributions of relaxed and unrelaxed haloes in the parameter space. Figure \ref{fig:obslims} depicts the $d_\text{off}-\Delta m_{12}$ relation for both relaxed and unrelaxed haloes. We choose an 80\% relaxation fraction in the combination of these two parameters to be a sufficient limit for relaxation. The dashed black lines in the figure indicate the regions where this applies. The horizontal line bounds the magnitude gap by $\Delta m_{12} > 1.6~\text{mag}$ and the vertical line the BHG offset by $d_\text{off} < 0.05~R_{200}$. These limits are chosen to align well with the distribution of the number density of relaxed haloes at different parameter values (shown with histograms in \mbox{Fig. \ref{fig:obslims})}. The haloes that fall within these limits are hereafter referred to as the proxy sample. We calculate the efficiency (precision) and completeness (recall) of this sample. Efficiency is defined as the fraction of relaxed haloes within the selection limits, relative to the total number of haloes within those limits. This is equivalent to the relaxed fraction, yielding an efficiency of 80\%. Completeness shows the fraction of relaxed haloes that fall within the selection limits relative to the total number of relaxed haloes in the data. We find the completeness to be 70\%.

The fraction of relaxed haloes at each parameter-pair grid point can also be inferred directly from this data. It can be seen in Fig.~\ref{fig:relfrac_from_data_3D2D} (left), where at each grid point the fraction of relaxed haloes is shown with a colour: red shades indicating relaxed and blue indicating predominantly unrelaxed regions. The dashed black lines indicate the found observable proxy limits as described above. In the right panel of the figure, we show the relaxation fraction when the distance is evaluated as an $xy$-plane projected distance $d_\text{off, proj}$ to test the applicability of this limit to observations. It is evident that, even though the general relaxed fraction is lower with projected offsets, the limit successfully constrains the relaxed sample from the unrelaxed one. Additionally, we can see that the relaxation probability shown in Fig.~\ref{fig:relfrac_combined_multip} is stricter than when using data directly. This indicates that the mean curves for each parameter might slightly underestimate the general trends; therefore, we use the limits directly derived from the $d_\text{off}-\Delta m_{12}$ parameter space.

The limits we have set are roughly within the same ranges as those of previous observational studies that have considered similar proxies for massive clusters. We find that our magnitude gap limit is more restrictive: for instance, \citet[relaxed fraction 50--70\%]{lopes_optical_2018} found a best break value between relaxed and unrelaxed systems at \mbox{$\Delta m_{12} > 1$~mag}, and \citet[relaxed fraction 19\%]{casas_optical_2024} found relaxed systems to have magnitude gap \mbox{values} of $\Delta m_{14} > 1.25$~mag. Our BHG offset values are generally less strict compared to results obtained from X-ray offsets: \citet[relaxed fraction 65\%]{lavoie_xxl_2016} restricted their sample by \mbox{$d_{1X} < 0.05~R_{500}$} to account for relaxed clusters, \citet{lopes_optical_2018} found a limiting value of \mbox{$d_{1X} < 0.01~R_{500}$}, and \citet{casas_optical_2024} suggested a constraint of \mbox{$d_{1X} < 15$~kpc}. This shows that using multiple partially dependent parameters allows one to vary the limits slightly while still maintaining a high relaxation fraction. Given that all three studies have smaller sample sizes than this work and use only one criterion for determining relaxation, we consider the limits derived from this analysis to be more robust and complete for estimating the halo's dynamical state. 

\begin{figure}
    \centering
    \includegraphics[width=1.\linewidth]{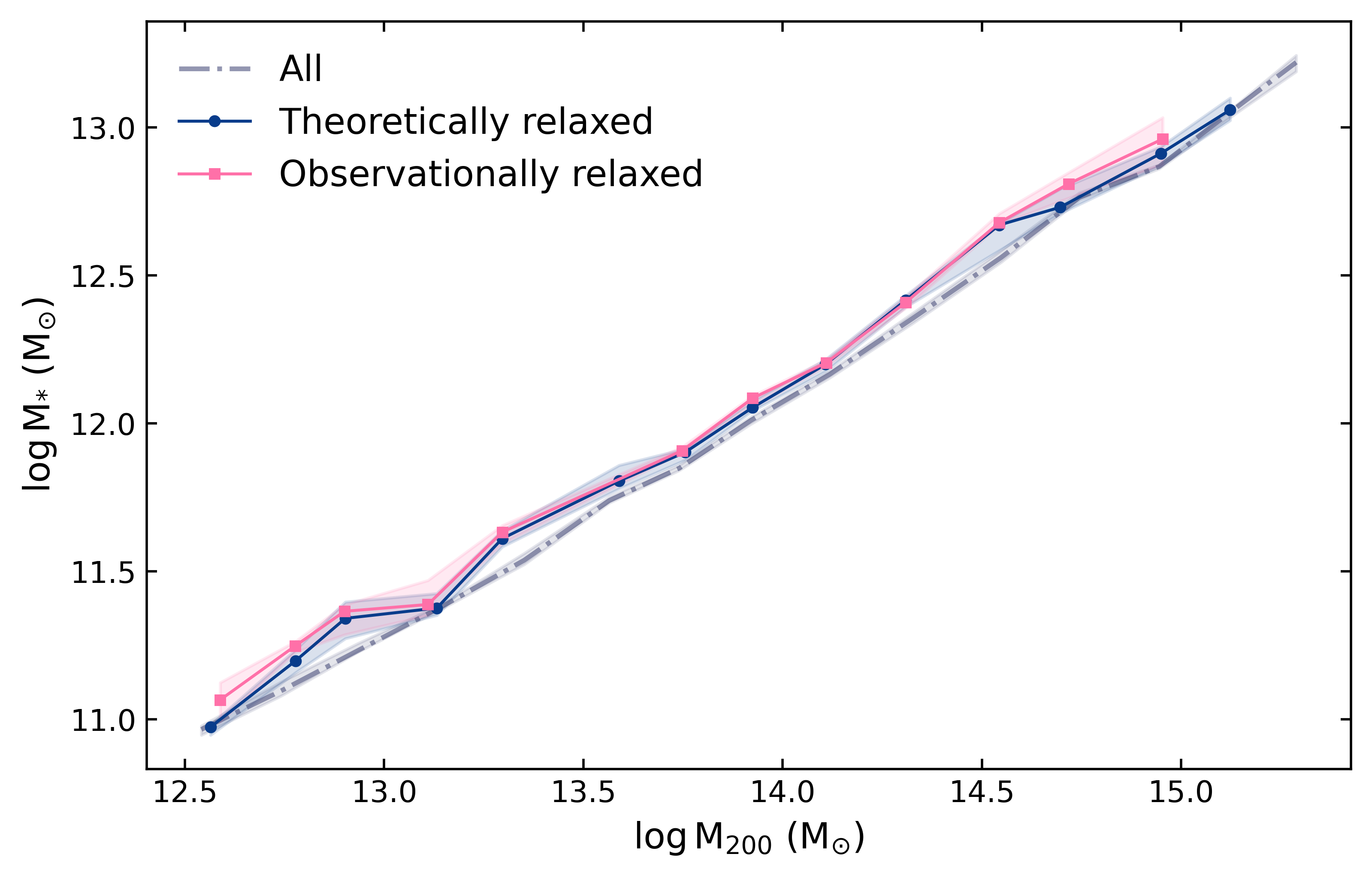}
    \caption{Median BHG stellar mass as a function of the virial mass of the host halo at $z=0$. The haloes identified as relaxed using theoretical criteria are shown in dark blue, and those selected using observational criteria are plotted in pink. The dot-dashed line represents the full halo sample. The two relaxed samples are consistent within the 68\% bootstrapping confidence intervals.}
    \label{fig:MstarM200_rescorr_relaxed_oblims}
\end{figure}

We investigated the relation between the BHG stellar mass and halo virial mass for the relaxed and total samples. This is important for assessing whether, and to what extent, the typically considered BHG--halo scaling relation is biased by the dynamical state, and if negating it affects the estimated \mbox{values.} 
This relation is displayed in Fig. \ref{fig:MstarM200_rescorr_relaxed_oblims}, where the median values of both theoretically and observationally determined relaxed systems are plotted alongside the whole sample. It can be seen that the theoretically and observationally relaxed haloes overlap within the 68\% bootstrapping confidence intervals and that the BHGs of relaxed haloes seem to be slightly more massive, especially in the low halo mass end. For a clearer overview of how BHG mass depends on its host halo mass, we can approximate a linear relation between them. 
Using linear regression, we find that the theoretically relaxed haloes follow a relation: $\log\text{M}_* = (0.809 \pm 0.008)~\log\text{M}_{200}+(0.799 \pm 0.113)$ and the observational proxy haloes are within the same range: $\log\text{M}_* = (0.804 \pm 0.008)~\log\text{M}_{200} + (0.902 \pm 0.106)$. The general sample (including both relaxed and unrelaxed haloes) follows $\log\text{M}_* = (0.850 \pm 0.004)~\log\text{M}_{200} + (0.190 \pm 0.051)$
Here, both the halo and BHG masses are expressed in units of $\text{M}_\odot$. When assessing the BHG masses from the halo virial masses, the trends show that until around $\text M_{200} = 10^{15}~\text M_\odot$, the BHGs of relaxed systems are more massive. To quantify whether this larger mass is a significant result, we compare the medians of BHG stellar mass distributions. The median value for all haloes is $10^{11.80}~\text M_\odot$, for theoretically relaxed haloes it is $10^{11.96}~\text M_\odot$, and $10^{11.93}~\text M_\odot$ for the observational proxy. This indicates that the relaxed haloes are about $1.3-1.5$ times more massive. 
To estimate the uncertainties of these medians, we use non-parametric bootstrapping with a 99.7\% confidence interval. The results show that the two relaxed subsamples (theoretical and observable) yield the same intervals, whereas using the whole halo sample yields consistently lower median values, confirming that the BHG stellar masses are larger in relaxed systems.
This result is in accordance with the findings of \citet{zenteno_dynamical_2025}, who show that the BHGs in relaxed clusters (defined as $d_{1X} < 0.25~R_{500}$) are brighter than BHGs in unrelaxed and diverse clusters. Additionally, earlier studies on present-epoch clusters such as \citet{wen_substructure_2013, lauer_brightest_2014, wen_dependence_2015} reported similar trends. This underscores the importance of accounting for the dynamical state in direct mass estimations. Given that it is more straightforward to estimate the stellar masses of central galaxies than the virial masses of large haloes, we use the derived linear approximations to determine the limiting BHG stellar masses. For relaxed haloes with a virial mass $\text M_{200} \geq 10^{12.5} \text{M}_\odot$, its BHG stellar mass is $\text{M}_* \geq 10^{10.9}~\text{M}_\odot \approx 7.9 \cdot 10^{10}\text{M}_\odot$.

\subsection{Redshift evolution} \label{redshift_evol}

\begin{figure}
    \centering
    \includegraphics[width=1.\linewidth]{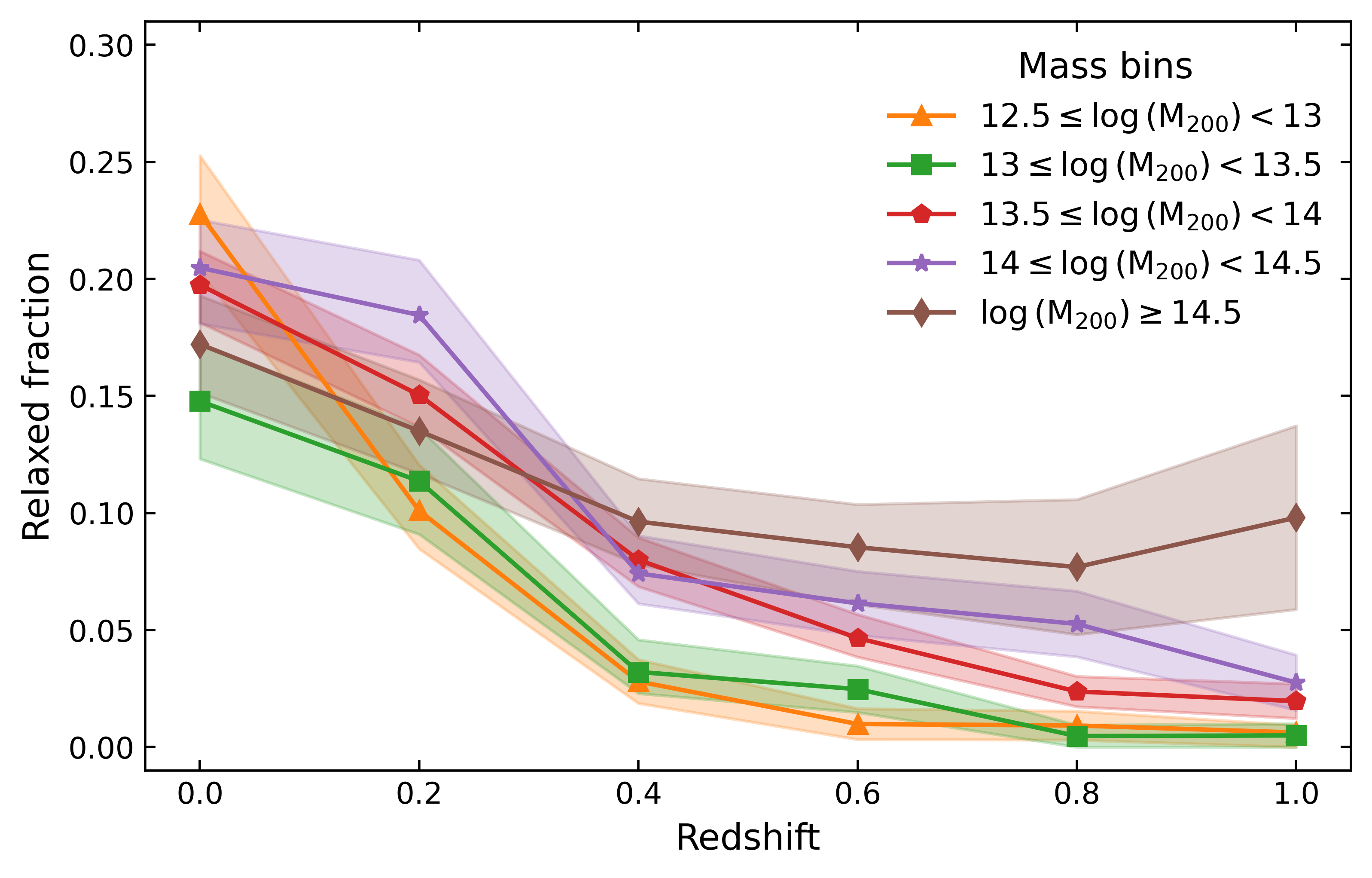}
    \caption{Halo relaxation fraction as a function of redshift. Relaxation is defined by the theoretical criteria defined in Sect. \ref{analysis_halo_relax}. The error bands indicate a 68\% bootstrapping confidence interval.}
    \label{fig:relfrac_massbin}
\end{figure}

\begin{figure}
    \centering
    \includegraphics[width=1.\linewidth]{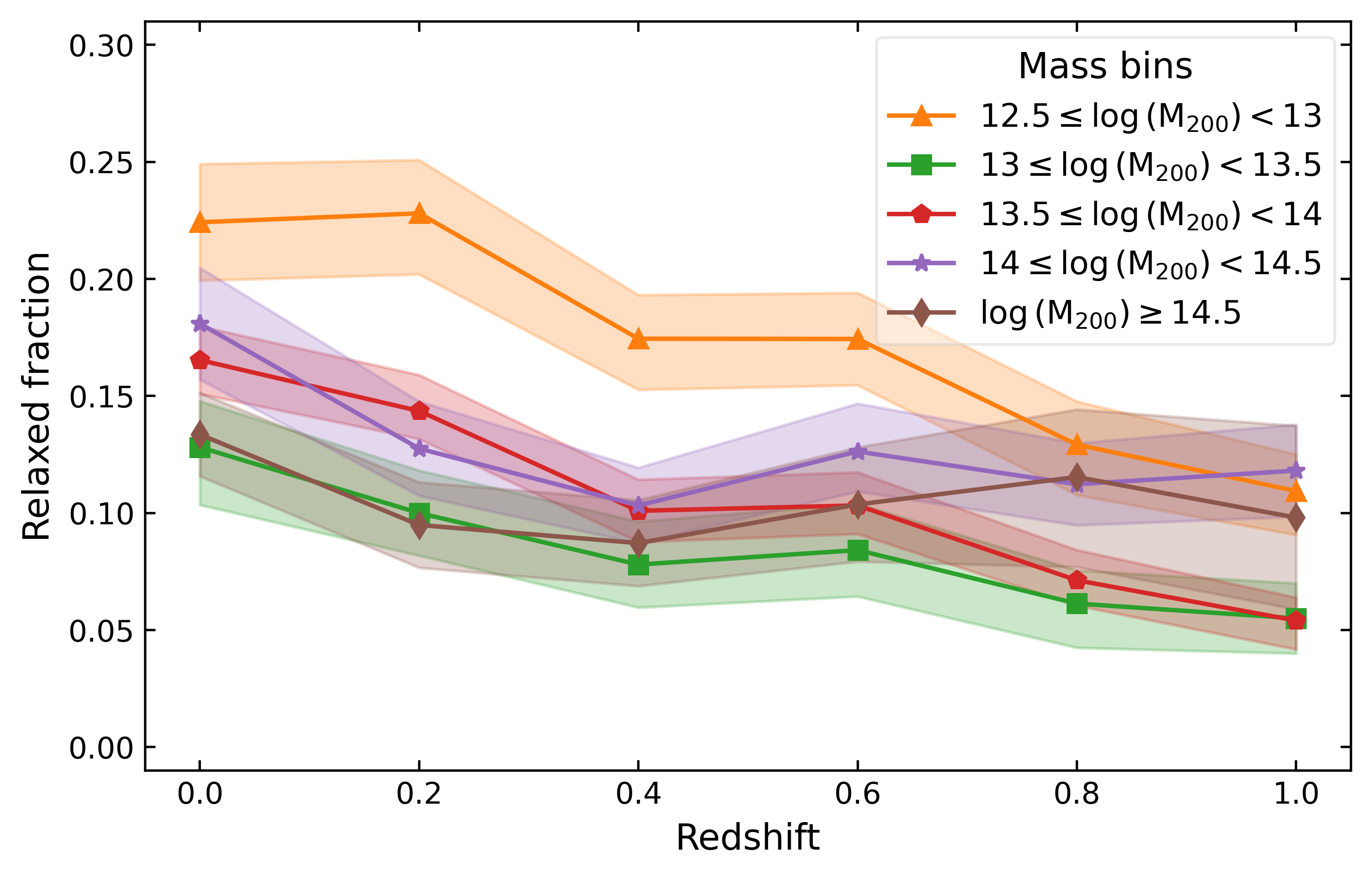}
    \caption{Halo relaxation fraction as a function of redshift when the observable constraints $d_\text{off} < 0.05~R_{200}$ and $\Delta m_{12} >1.6~\text{mag}$ are used as relaxation criteria. The error bands indicate a 68\% bootstrapping confidence interval.}
    \label{fig:relax_z_obslims}
\end{figure}

We examined how the total fraction of relaxed haloes evolves with redshift. The calculations were done in the same mass bins as previously at $z=0$. The evolution covers the range $z=0$ to $z=1$ with a step of $0.2$. The relaxation of haloes is defined in two ways: based on the theoretical criteria (Fig. \ref{fig:relfrac_massbin}) and based on the observable proxies chosen in Sect. \ref{obs_prox_rel} (Fig. \ref{fig:relax_z_obslims}).
The general trend suggests that the closer we approach the present time, the larger the fraction becomes. For theoretically relaxed haloes shown in Fig. \ref{fig:relfrac_massbin}, there is a notable increase beginning at redshift $z=0.4$, indicating that relaxation speeds up. This value has been associated with multiple occurrences during group and cluster evolution and the change in their dynamical state. 
For instance, \citet{mann_x-ray-optical_2012} found that the cluster merger rate drops significantly at redshifts below $z=0.4$, meaning that the amount of interactions that would disrupt relaxation decreases, giving haloes the opportunity to reach relaxation. In parallel, the analysis of cluster progenitors by \citet[see Fig. 2 in their work]{chiang_ancient_2013} showed that clusters reach their effective radii at around $z=0.4$, possibly indicating that the systems start to become more stable at this point in their evolution.

The relaxation fraction in the most massive bins, on the other hand, increases far more slowly. This indicates that these very massive systems have formed much later through mergers, and many of them have not yet had time to reach dynamical equilibrium or become virialised. The delayed relaxation of massive systems is supported by the findings of \citet{amoura_cluster_2021}. 
They showed that haloes with a current mass of roughly $\text{M} \gtrsim 10^{14}~\text{M}_\odot$ reached half of their mass around $z=0.5$, meaning that at least one major merger event would still occur before $z=0$. This mass evolution can also be traced by their BHG: \citet{de_lucia_hierarchical_2007} reported that the brightest galaxies of massive clusters also reach half of their final mass at $z\sim 0.5$, which again supports the idea that the BHG is a good proxy for estimating its host halo properties throughout their co-evolution.

When defining relaxation based on the observable proxies, the redshift evolution of relaxation changes considerably. While the fractions closely match in both approaches at $z=0$, they differ significantly at earlier evolution steps. The fraction is consistently higher at halo masses $12.5 \leq \log \text{M}_{200} < 13$, indicating that at lower masses (and higher redshifts), we have higher contamination by unrelaxed haloes in the data. 
For all mass bins in Fig.~\ref{fig:relax_z_obslims}, there is a small dip in relaxation fraction at $z=0.4$, which possibly can be related to the discussion above regarding halo evolution. On the other hand, there seems to be no evident reason why there would be an increase in relaxation at $z=0.6$. Given that, within the bootstrapping error limits, the fluctuation peak is negligible, we assume that it is a by-product of the current relaxation definition. This means that these observable proxies may have an upper limit on how far they can be applied, and that these observables likely also evolve with redshift.

\begin{figure}
    \centering
    \includegraphics[width=1.\linewidth]{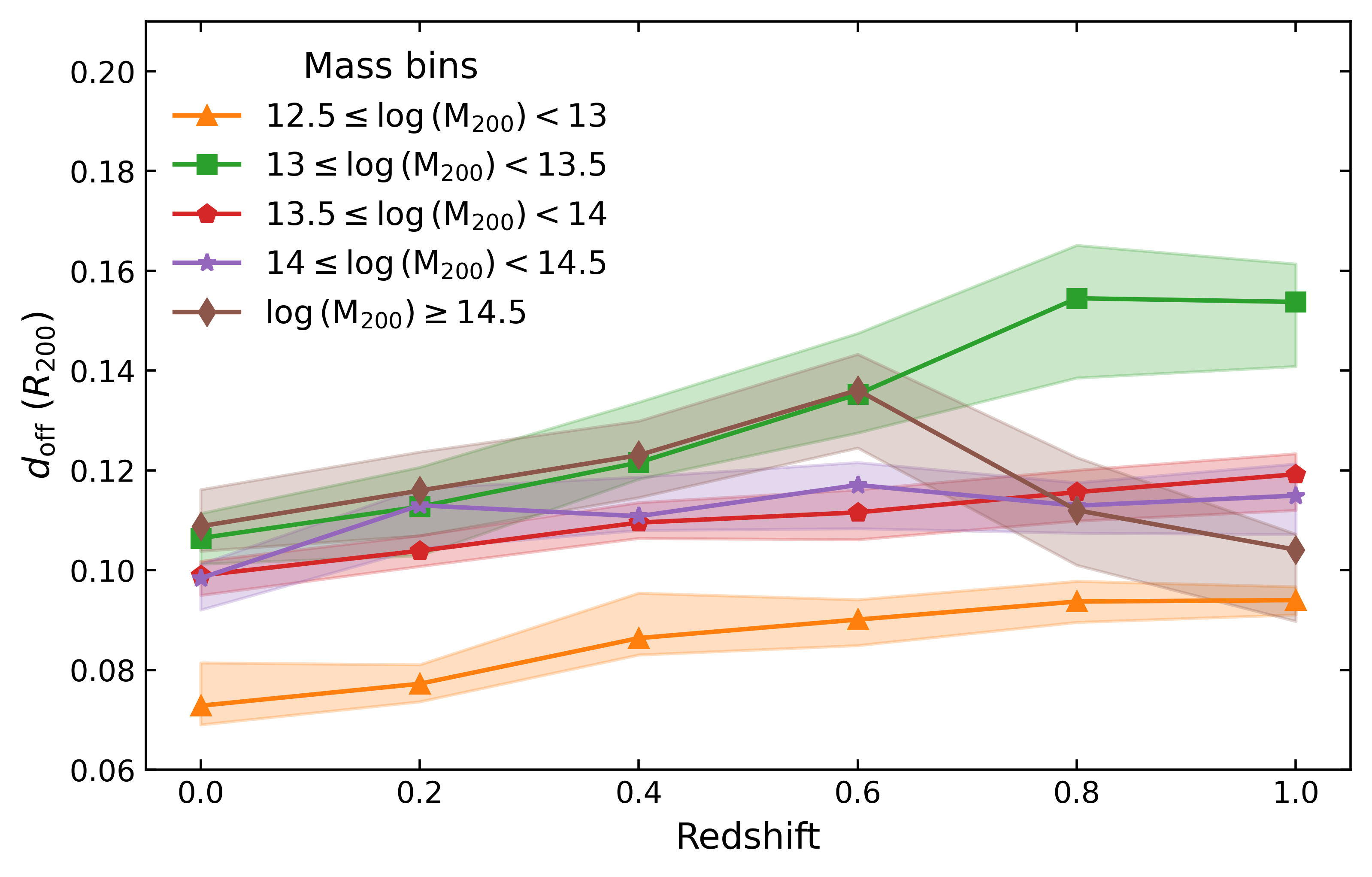}
    \caption{Median values of the BHG offset $d_\text{off}$ from the cluster centre at each redshift for the full halo sample. The error bands indicate 68\% bootstrapping confidence intervals.}
    \label{fig:doff_R200_massbin}
\end{figure}

\begin{figure}
    \centering
    \includegraphics[width=1.\linewidth]{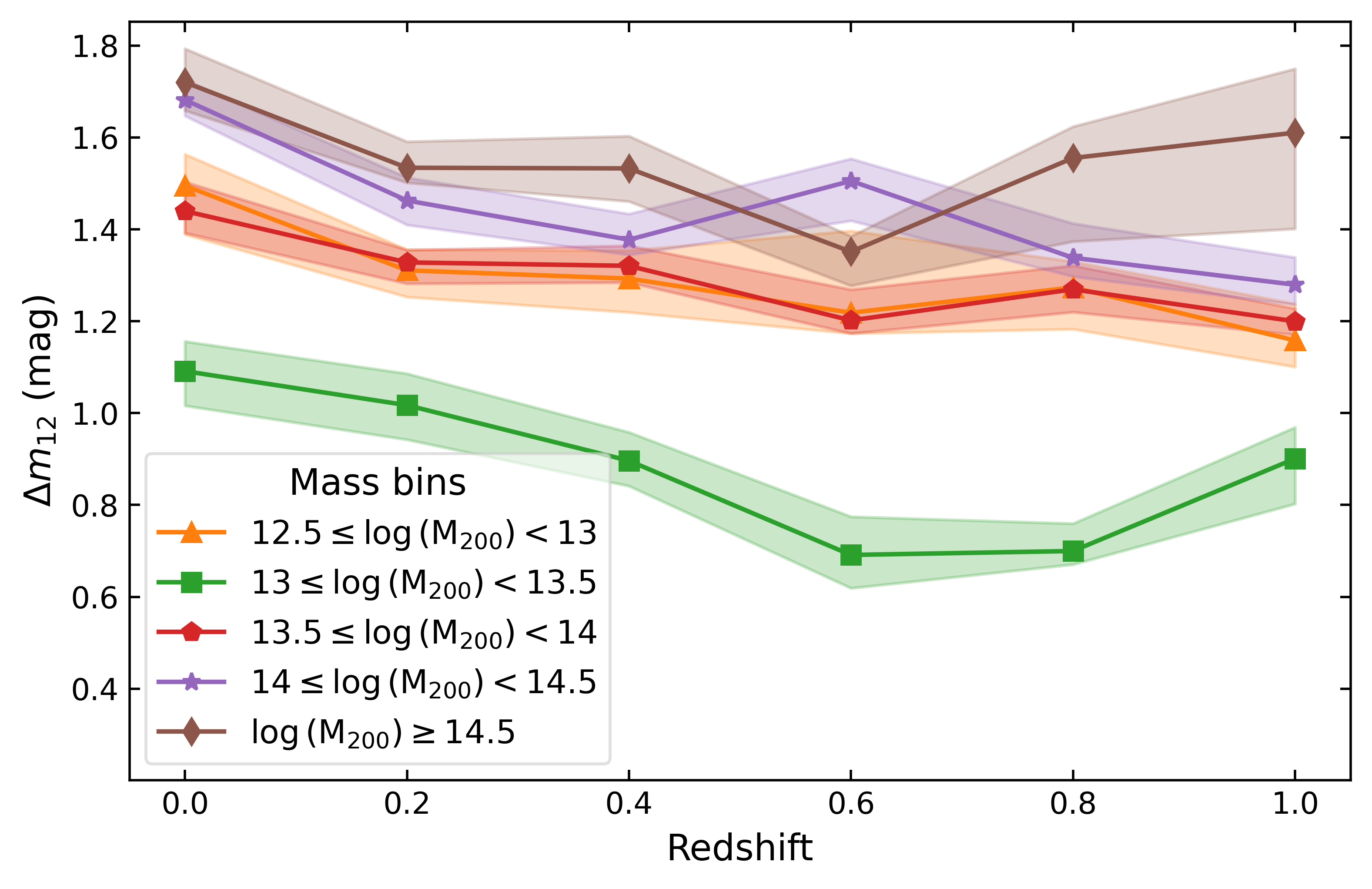}
    \caption{Median values of the magnitude gaps $\Delta m_{12}$ at different redshifts for the full halo sample. Error bands indicate 68\% bootstrapping confidence intervals.}
    \label{fig:m12_massbin}
\end{figure}

\begin{figure}
    \centering
    \includegraphics[width=1.\linewidth]{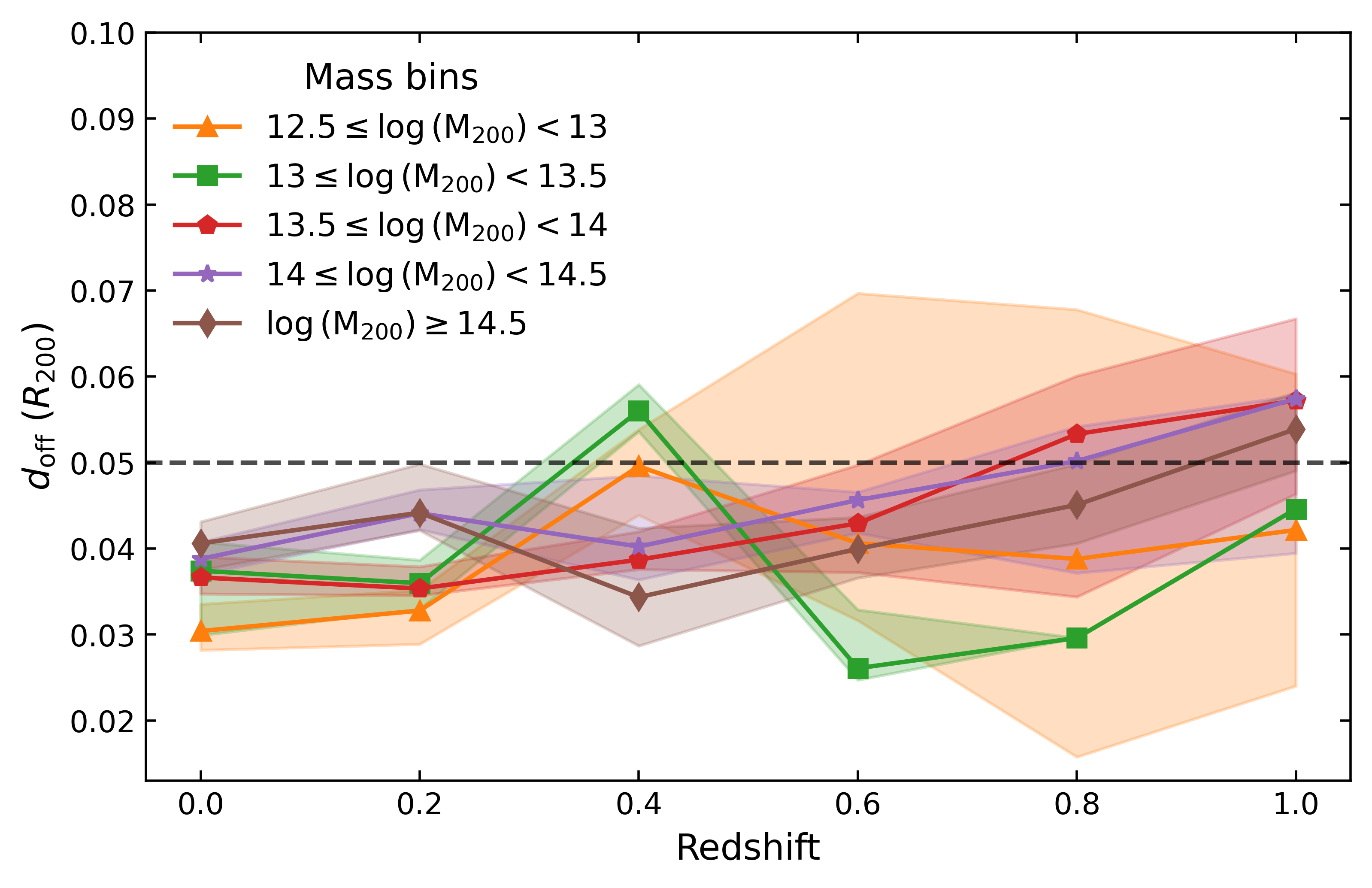}
    \caption{Median values of BHG offsets $d_\text{off}$ at different redshifts for (theoretically defined) relaxed haloes. The horizontal black line shows the offset limit for relaxed systems. Error bands display 68\% bootstrapping confidence intervals.}
    \label{fig:doff_R200_massbin_relaxed}
\end{figure}

\begin{figure}
    \centering
    \includegraphics[width=1.\linewidth]{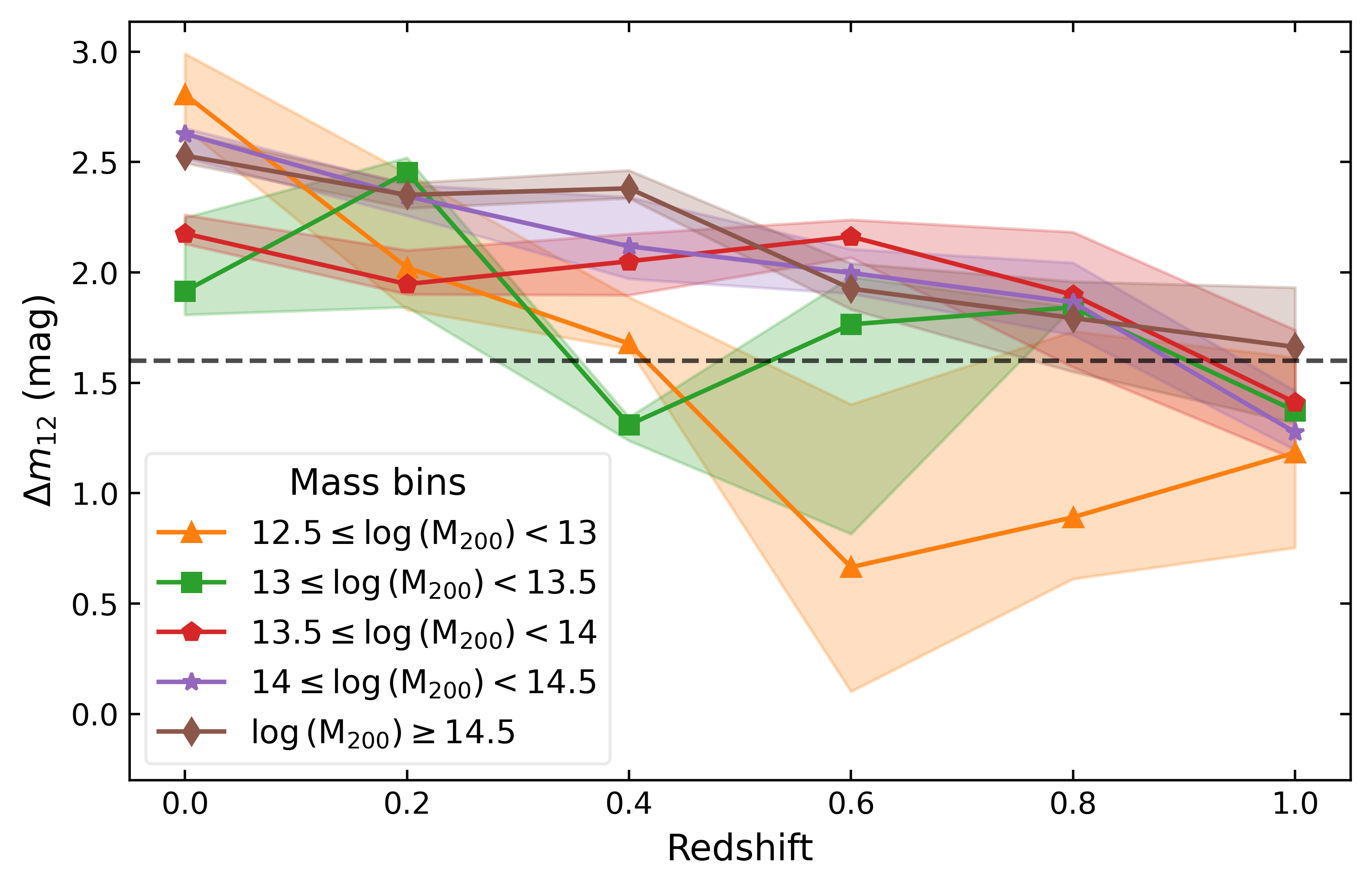}
    \caption{Median values of the magnitude gaps $\Delta m_{12}$ at different redshifts for (theoretically defined) relaxed haloes. The horizontal black line shows the magnitude gap limit for relaxed systems. Error bands display 68\% bootstrapping confidence intervals.}
    \label{fig:m12_massbin_relaxed}
\end{figure}

We trace the evolution of the BHG (and SBHG) properties used to define relaxation observationally. We analyse the evolution of the BHG offset and the magnitude gap for the entire sample and for fully relaxed haloes (defined by the theoretical criteria). The median values of the observable parameters ($d_\text{off},~\Delta m_{12}$) at each redshift for the full sample are shown in Figs. \ref{fig:doff_R200_massbin} and \ref{fig:m12_massbin}. The general trend shows an increase in $d_\text{off}$ with increasing redshift in nearly all mass bins. The only outlier in this case is the most massive halo range, where the median BHG distance increases until $z=0.6$. We assume this is related to major merger events that form these massive systems. However, to confirm this, the systems' evolution would need to be traced and analysed throughout their entire formation history, which is out of scope for this work. The relation for the median magnitude gap shows no clear redshift dependence in any mass bin, as the values fluctuate substantially throughout halo evolution.

The same relations are presented for fully relaxed haloes in Figs.~\ref{fig:doff_R200_massbin_relaxed} and \ref{fig:m12_massbin_relaxed}. 
Here we observe a slightly clearer systematic relation with redshift. Given that there are only a few low-mass relaxed systems at high redshifts, the bootstrapping errors are quite wide at the higher-z end. Overall, excluding some fluctuations, we note that BHG offsets decrease and the magnitude gaps rapidly increase with decreasing redshift. Additionally, the median offset values for relaxed haloes are much lower than for the general sample, already at $z=1$, and the magnitude gaps are notably larger. When accounting for redshift evolution of these properties for relaxed haloes, we suggest that the reliable upper limit for using these observable proxies is around $z\sim0.2$, as at this redshift, the median values of the observables are within the set limits (horizontal lines, $d_\text{off} < 0.05~R_{200}$, $\Delta m_{12} > 1.6~\text{mag}$). After this point, in some mass bins, the limits do not reliably reflect the general trends of the relaxed haloes. 

\subsection{The halo mass function}

The halo mass function describes the halo distribution, that is, typically the number of haloes at different mass values. As was discussed above, the HMF of relaxed haloes is a useful tool for constraining cosmological parameters. In this section we examine how halo dynamical masses and their mass function differ when considering or ignoring halo relaxation. Relaxation is defined by theoretical criteria shown in Sect.~\ref{analysis_halo_relax}. We analyse how these relaxed halo mass functions compare with those derived from the proxy sample defined in Sect. \ref{obs_prox_rel}. We evaluate the mass functions separately for the TNG100-1 and TNG300-1 simulation runs. TNG-Cluster is excluded as its most massive haloes are not representative of the full simulation volume.  

To recover the halo mass function from the simulation runs, we estimate halo masses using only the galaxies identified as halo members after the data reductions, rather than the true simulation halo mass $\text{M}_{200}$. This procedure accounts for selection effects that can affect mass estimates in observations. The dynamical mass is calculated with the virial mass relation,
\begin{equation}
    \quad M_\text{dyn} = \dfrac{R\sigma
    ^2}{G},
\end{equation}
 where $G$ is the gravitational constant, $R$ is the virial radius ($R_{200}$) and the velocity dispersion is expressed as 
\begin{equation}
    \sigma = \sqrt{\dfrac{1}{N-1}\sum_i^N(v_i-\bar{v})^2},
\end{equation}
where $v_i$ is the total (scalar) velocity of each galaxy $i$ in its host cluster, $\bar{v}$ is the average velocity of galaxies in the halo, and $N$ is the total number of galaxies within the halo's virial radius. Here, we use all three velocity components for the calculations.

\begin{figure}
    \centering
    \includegraphics[width=1.\linewidth]{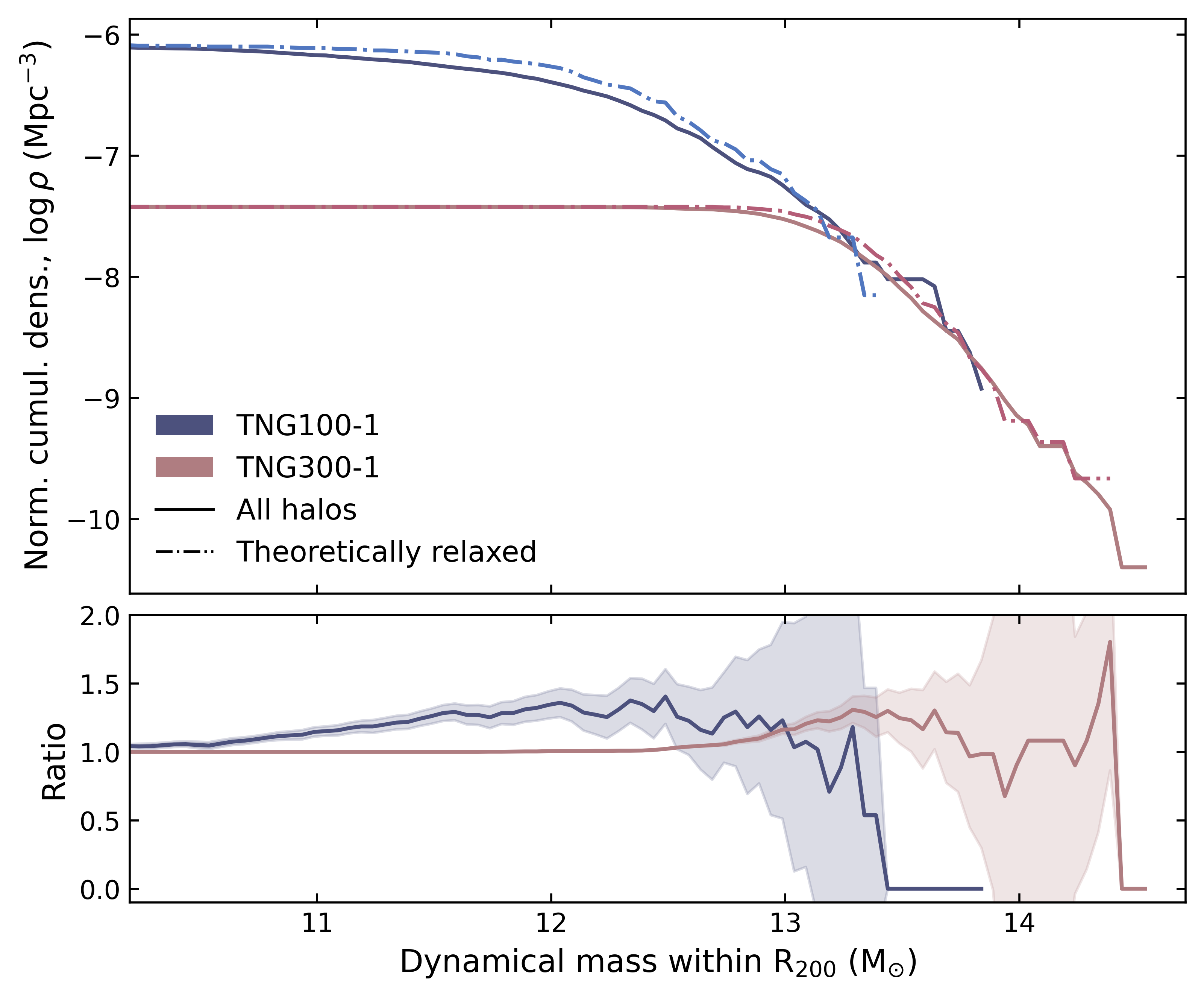}
    \caption{\textit{Top}: Halo mass function for each simulation run. Solid lines represent all the data, and dot-dashed lines represent dynamically relaxed haloes. \textit{Bottom}: Ratio of relaxed haloes over all haloes. The errors are 68\% bootstrapping confidence intervals.}
    \label{fig:massfuncs}
\end{figure}

\begin{figure}
    \centering
    \includegraphics[width=1.\linewidth]{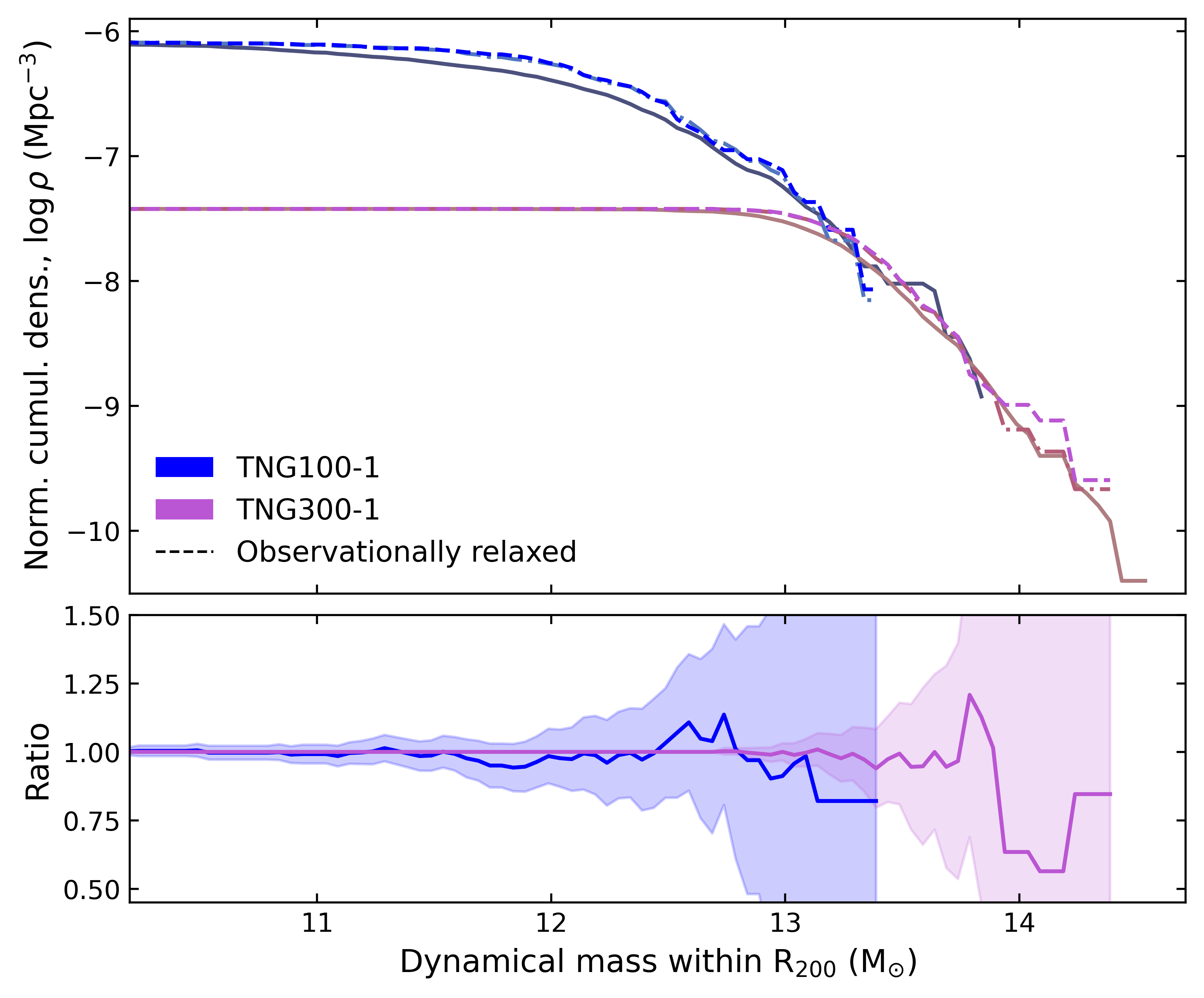}
    \caption{\textit{Top}: Halo mass functions obtained using observable proxies for the relaxation plotted with a dashed line over the mass functions shown in Fig. \ref{fig:massfuncs}. \textit{Bottom}: Ratio of relaxed haloes over the proxy sample. The errors are 68\% bootstrapping confidence intervals. It is evident that the observable proxies are sufficient for estimating relaxation, and the mass functions overlap well.}
    \label{fig:massfunc_obs}
\end{figure}

The cumulative mass functions are shown in Fig. \ref{fig:massfuncs}, where each simulation run is displayed in a different colour. The mass functions are scaled by the size of the datasets used for the calculations. It can be seen that there is a difference in the knee of the mass functions, which is especially evident in TNG100-1. To confirm the difference in distributions, non-parametric statistical tests, such as the Mann–Whitney U test and the Kolmogorov-Smirnov test, were performed on the mass functions. We chose a confidence level of 95\%, meaning that the null hypothesis of the data being drawn from the same distribution is rejected if the p-value is less than 0.05. The results consistently showed that the underlying distributions differed, that is, the null hypothesis is being rejected. This is also visually confirmed in the bottom panel of the figure, where we display the ratio of relaxed haloes to all haloes. The error bands are expressed as 68\% bootstrapping confidence intervals. 
The magnitude of the relation is mass-dependent but remains positive throughout the mass range. The bias peaks at about 40\% (when including errors, at minimum 20\%) around mass $10^{13}~\text{M}_\odot$ for TNG100-1 and just below $10^{14}~\text{M}_\odot$ for TNG300-1.

Figure \ref{fig:massfunc_obs} overlays the mass functions of the proxy sample
haloes on the previous two. The proxy sample was defined above as \mbox{$\Delta m_{12} > 1.6~\text{mag}$}, \mbox{$d_\text{off} < 0.05~R_{200}$}.
It can be seen that the reconstructed mass function overlaps with the relaxed halo selection. These distributions were again compared using the same non-parametric tests to confirm that the visual conclusions are statistically consistent. The outcome indicates that data with 80\% efficiency and 70\% completeness are sufficient to replicate the mass functions of relaxed haloes, making the constraints for the proxy sample a suitable tool for estimating the dynamical state of groups and clusters using low-cost observable properties.

\section{Discussion and conclusion} \label{summary} 

Our work provides a general framework for evaluating the relaxation of galaxy groups and clusters using simple observable properties of their brightest galaxies: (1) the BHG offset from the halo centre $d_\text{off} < 0.05~R_{200}$, and (2) the magnitude gap between the two brightest galaxies $\Delta m_{12} > 1.6~\text{mag}$. 
The distance between the two brightest galaxies $d_{12}$ is not obviously correlated with the dynamical state of the halo, and the BHG distance from the luminosity centre ($d_\text{lum}$) provided a strong correlation only for haloes with masses $\text{M}_{200} \leq 10^{13.5}~\text{M}_\odot$. 
The chosen criteria were applied to haloes with a virial mass greater than $10^{12.5}~\text{M}_\odot$. 
When the virial mass is also to be approximated as a more easily observationally obtainable property, we suggest to refer to the scaling relation of the BHG stellar mass and halo virial mass (Fig. \ref{fig:MstarM200_rescorr_relaxed_oblims}). Using a linear approximation between these masses, we found that the halo virial mass limit $\text{M}_{200} \geq 10^{12.5}~\text{M}_\odot$ roughly corresponds to a BHG stellar mass value of $\text{M}_*\geq10^{10.9}~\text{M}_\odot$ for relaxed haloes. We note that this scaling relation is dependent on the halo dynamical state: the BHGs of relaxed haloes are typically more massive than those in unrelaxed haloes, especially in the lower halo mass end. The relations intersect at $\text{M}_{200} \sim 10^{15}\text{M}_\odot$.

Even though the constraints for the offsets and magnitude gaps were found using local objects ($z=0$), the redshift evolution of these properties (Figs.  \ref{fig:doff_R200_massbin_relaxed}, \ref{fig:m12_massbin_relaxed}) suggests that they can be reliably applied to redshifts up to $z\sim 0.2$, where the median offsets and magnitude gaps of relaxed haloes are well within the limits. At higher redshifts, the constraints become less reliable. The decrease in reliability can be related to the redshift evolution of the halo relaxation fraction (Fig.~\ref{fig:relfrac_massbin}), where we noted a drop-off in relaxation at $z=0.4$. This implies that when the relaxation fraction is low, the few haloes that are relaxed in the sample lack sufficiently distinctive properties to tell them apart from the rest of the sample. Details of higher-redshift systems need to be analysed thoroughly to reach any conclusions about more distant systems. Based on the current results, we recommend using these limits up to $ z\sim0.2$ to obtain reliable results.

The successful replication of the relaxed HMF supports the use of these criteria to identify the dynamical state. Nevertheless, it remains essential to apply them to observational data of galaxy clusters and groups to confirm their validity and to quantify the agreement between theoretically derived parameters and real systems. For one, even as the resolution of large-scale simulations has improved, it might still be insufficient to reliably resolve the lowest-mass systems and fully capture their dynamical state. As a result, our identified reliability limit of $10^{12.5}~\text{M}_\odot$ might still be affected by resolution effects that prevent smaller systems from being properly characterised. We speculate that when a more complete sample of the less massive end of galaxy systems is provided, these systems might also show stronger signs of relaxation. To account for this, simulations and observations focused specifically on lower-mass galaxy groups must be considered. As for our own efforts, we will in future work study the relaxed systems in observations using the latest Galaxy and Mass Assembly (GAMA) group catalogue, since this survey has the most complete sampling of galaxy groups to date. We will analyse the relaxation of these observed systems to further confirm the applicability of the relaxation proxies we found. In addition, recently launched survey programmes, such as the 4-metre Multi-Object Spectroscopic Telescope (4MOST) Wide Area Vista Extragalactic Survey (WAVES) \citep{2019Msngr.175...46D} and the 4MOST Hemisphere Survey (4HS) \citep{2023Msngr.190...46T}, will further improve the coverage of low-mass haloes, reaching masses comparable to those of the Magellanic Clouds. This provides an opportunity for analysing the dynamical state of galaxy systems throughout the full mass continuum, paving the way for a better understanding of how these systems form and evolve within their cosmic web environments.

The total fraction of fully relaxed haloes at $z=0$ was $15-23\%$, placing it at the lower end of observational estimates ($13-70\%$), which vary depending on the method. Excluding observational biases (e.g. cool-core selection effects in X-ray observations), we speculate that this discrepancy primarily originates from the definition of relaxation. Many studies, including \citet{lavoie_xxl_2016, seppi_offset_2023, zenteno_dynamical_2025}, relied on the BHG–X-ray centre offset alone as a relaxation criterion. While this metric provides a useful initial proxy for the dynamical state, it has notable limitations. Firstly, a single-parameter definition can misclassify haloes when unrelaxed systems transiently satisfy the criterion, thereby increasing contamination. Secondly, achieving high precision with a single metric requires strict thresholds, which reduces completeness by excluding relaxed systems that lie near the chosen threshold. A more robust approach is to combine multiple (a minimum of two) complementary but independent indicators. As demonstrated here, individual criteria impose tighter constraints in isolation, whereas their joint application allows us to use more flexible thresholds while improving the sample purity and reliability. 

Another limitation of the X-ray centre proxy is that it does not perfectly trace the true halo centre and can be significantly offset from the potential minimum. Using simulations, \citet{cui_how_2016} showed that the X-ray centre is often more displaced from the potential minimum than the BHG, indicating that the centre of X-ray emission is not a reliable tracer of the gravitational potential well. In addition, strong X-ray emission is predominantly observed in massive clusters, biasing such samples towards high-mass cool-core systems. We found similar behaviours for other baryonic tracers: the centre of light is also typically more offset from the potential minimum than the BHG, although it does perform better in lower-mass haloes (see Fig. \ref{fig:relaxed_dlum}), and seems to perform at least as well as the X-ray centre estimate. Replacing the true potential well offset ($d_\text{off}$) with the centre-of-light distance criterion $d_\text{lum} < 0.05~R_{200}$ leads to a noticeable decline in the predictive performance. In combination with the magnitude gap, the efficiency drops to 40\% and the completeness to 59\%, compared to 80\% and 70\%, respectively, when $d_\text{off}$ is used. This suggests that baryonic components, which are subject to complex dynamical processes, fail to consistently trace the gravitational potential throughout the whole mass continuum. An accurate and more general tracer of the halo potential minimum is therefore still needed. One of these possible tracers might be the ICL. In recent work, \citet[and references therein]{fernandez_intracluster_2026} suggested the use of the ICL to approximate the shape of cluster DM haloes. As the ICL is shown to follow the gravitational potential more closely than satellite galaxies, it might be a much more precise proxy for the DM halo shape and orientation, potentially enabling an accurate estimate of the location of the halo centre. 
Upcoming survey data will provide further insight into the feasibility of this approach and whether better instrumentation also makes it applicable to galaxy groups.

\begin{acknowledgements}
    We thank the referee for the feedback, valuable comments, and suggestions, which helped us to improve the paper.
    This work was funded by the Estonian Ministry of Education and Research (grant TK202), Estonian Research Council grants (PRG3034 and PRG2172), the European Union's Horizon Europe research and innovation programme (EXCOSM, grant No. 101159513) and the Vilho, Yrjö and Kalle Väisälä Foundation.
\end{acknowledgements}

\bibliographystyle{aa}
\bibliography{references}

\end{document}